\documentclass[10pt]{iopart}
\usepackage{graphicx}
\usepackage{dcolumn}
\usepackage{array}
\usepackage{iopams} 
\usepackage{bm}

\begin{document}

\title{Semirelativistic Hamiltonians and the auxiliary field method}

\author{Bernard Silvestre-Brac$^1$, Claude Semay$^2$, Fabien Buisseret$^2$}

\address{$^1$ LPSC Universit\'{e} Joseph Fourier, Grenoble 1,
CNRS/IN2P3, Institut Polytechnique de Grenoble, 
Avenue des Martyrs 53, F-38026 Grenoble-Cedex, France}
\address{$^2$ Groupe de Physique Nucl\'{e}aire Th\'{e}orique, Universit\'{e}
de Mons, Acad\'{e}mie universitaire Wallonie-Bruxelles, Place du Parc 20,
B-7000 Mons, Belgium}
\eads{\mailto{silvestre@lpsc.in2p3.fr}, \mailto{claude.semay@umh.ac.be}, 
\mailto{fabien.buisseret@umh.ac.be}} 

\date{\today}

\begin{abstract}
Approximate analytical closed energy formulas for semirelativistic Hamiltonians of the form $\sigma\sqrt{\bm p^{\, 2}+m^2}+V(r)$ are obtained within the framework of the auxiliary field method. This method, which is equivalent to the envelope theory, has been recently proposed as a powerful tool to get approximate analytical solutions of the Schr\"{o}dinger equation. Various shapes for the potential $V(r)$ are investigated: power-law, funnel, square root, and Yukawa. A comparison with the exact results is discussed in detail.
\end{abstract}

\maketitle

\section{Introduction}

Since the beginning of quantum mechanics, there has been a considerable amount of works devoted to the computation of analytical solutions of the Schr\"{o}dinger equation, especially in bound states problems. Apart from its intrinsic mathematical interest, finding analytical formulas is always useful in physics, for example to obtain informations about the dependence of observables on the parameters of the Hamiltonian. This can be a great advantage when one tries to fit the parameters of a model to some experimental data. 

Many methods allowing to find approximate analytical solutions of the Schr\"{o}dinger equation for bound state problems have been extensively used in the literature: WKB method, variational method, \textit{etc.} (many solved problems can be found in \cite{flu} for example). Recently, we have proposed a new method to compute the bound states of a given Hamiltonian \cite{af,af2}. This method, called the auxiliary field method (AFM), is based on auxiliary fields -- also known as einbein fields \cite{af1} -- and can lead to approximate closed analytical results in many cases: One can quote central potentials of the form $a r^\lambda+br^\eta$ for special values of $\lambda$ and $\eta$ \cite{af2}, exponential potentials of type $-\alpha\, r^{\lambda}\, {\rm e}^{-\beta\, r}$ \cite{af3}, and square root potential \cite{hybri,afsqrt}. 

In standard Schr\"{o}dinger Hamiltonians, the kinetic term is of the form $\bm p^2/2m$. However, there exists another class of Hamiltonians in which the kinetic term is relativistic, typically
\begin{equation}\label{ham0}
	H=\sigma\sqrt{\bm p^2+m^2}+V(r),
\end{equation}
with $\sigma$ a positive number. Such Hamiltonians are usually called ``spinless Salpeter Hamiltonians". They are actually a well-defined approximation
of the Bethe-Salpeter formalism applied to the description of bound states within quantum field theory \cite{ssh}. The underlying assumptions is that the constituent particles interact instantaneously and propagate like free particles \cite{ssh2}. Moreover, the spin effects are not taken into account in this framework. For this reason one often speaks of ``semirelativistic Hamiltonians" to characterize the spinless Salpeter Hamiltonians. 

Because of the square root in the kinetic energy, it is no mystery that finding analytical eigenvalues for Hamiltonians like~(\ref{ham0}) is even more difficult than in the nonrelativistic, Schr\"{o}dinger, case. Such a task can however be fulfilled by using envelope theory (ET) \cite{env0}, which is a powerful method to get approximate analytical solutions of eigenequations, for various potentials \cite{Hall01,env,env3,env2,env4}. We have shown in a previous work that AFM and ET are, to a large extent, two  equivalent approaches \cite{afmenv}. That is why it is reasonable to suppose that, using the AFM, we will be able to get analytical formulas for the eigenvalues of spinless Salpeter-type Hamiltonians. That is the topic the present work is devoted to. 

Our paper is organized as follows. The general principles of the AFM are recalled in section~\ref{geneAFM}, and its application to spinless Salpeter Hamiltonians is discussed in section~\ref{srhamproc}. The method is applied to analytically solve several cases of interest in physics: Power-law potentials in sections~\ref{plpot} and \ref{secur}, square root potential in section~\ref{sqrtpot}, funnel potential in section~\ref{funnelpot}, and Yukawa potential in section~\ref{yukpot}. By comparing our formulas to numerical results, we then show in section~\ref{increase} that their accuracy can be considerably improved by performing some minor formal modifications. Finally we sum up our results in section~\ref{conclu}. Since the formulas we obtain are valid for one-particle systems or for two particles with the same mass, we give in~\ref{uneqma} some results in the unequal mass case, while some polynomial equations that are of interest in this work are presented in \ref{poleq}.

\section{The auxiliary field method}\label{geneAFM}
\subsection{General technique}

The auxiliary field method has been first presented in \cite{af} and extended in \cite{af2,af3}. In order for the present paper to be self-contained, we recall here the main points of the AFM and refer the reader to the aforementioned references for more details. 

The aim is to find an analytical approximate solution of the eigenequation $H \left|\Psi\right\rangle=E\left|\Psi\right\rangle$, with
\begin{equation}\label{inco}
	H=T(\bm p^{\, 2})+V(r).
\end{equation}
As usual, we work in natural units $\hbar=c=1$.
The AFM consists in the following procedure. We first assume that $H_A=T(\bm p^{\, 2})+\nu\, P(r)$, where $\nu$ is a real parameter, admits bound states with an analytical spectrum. The eigenequation  
\begin{equation}\label{analy}
	H_A \left|\Psi(\nu)\right\rangle=E_A(\nu)\left|\Psi(\nu)\right\rangle
\end{equation}
is thus analytically solvable, as well as the one associated with the Hamiltonian 
\begin{equation}\label{htdef}
\fl
\tilde H(\nu)=T(\bm p^{\, 2})+\tilde V(r,\nu) \quad \textrm{with} \quad \tilde V(r,\nu)=\nu\, P(r)+V\left(I(\nu)\right)-\nu\, P\left(I(\nu)\right).
\end{equation}
The function $I(x)$ is defined by 
\begin{equation}\label{rdef}
	I(x) =K^{-1}(x) \quad \textrm{and} \quad K(x)=\frac{V'(x)}{P'(x)},
\end{equation}
the prime denoting the derivative with respect to $x$.
Let us remark that $\tilde V(r,\nu)$ is of the form $C_1 P(r) + C_2$ where $C_1$ and $C_2$ are constants. These numbers are determined in order that $\tilde V(r)$ approximates at best the potential $V(r)$.  
Using~(\ref{analy}) we have
\begin{equation}\label{en1}
	E(\nu)=E_A(\nu)+V\left(I(\nu)\right)-\nu\, P\left(I(\nu)\right).
\end{equation}
The approximate eigenvalues and eigenstates are eventually given by $E(\nu_0)$ and $\left|\Psi_A(\nu_0)\right\rangle$ respectively, with $\nu_0$ such that the total energy~(\ref{en1}) is extremal, \textit{i.e.} satisfying
\begin{equation}\label{enmin}
\left.\partial_\nu E(\nu)\right|_{\nu=\nu_0}=0.
\end{equation}
The value of $\nu_0$ depends on the quantum numbers of the state considered. Once $\nu_0$ is known, the constants $C_1$ and $C_2$ can be computed. It is shown in \cite{afmenv} that $\tilde V(r_0,\nu_0)=V(r_0)$ and $\tilde V'(r_0,\nu_0)=V'(r_0)$, where $r_0$ is such that $\nu_0=K(r_0)$. 

The above procedure can be justified as follows. Let us assume that, instead of considering $\tilde V(r,\nu)$ in which $\nu$ is a constant, we consider $\tilde V(r,\hat \nu)$ in which $\hat \nu$ is an arbitrary function. A proper elimination of this auxiliary field $\hat \nu$ by the following constraint $\left.\delta_{\hat \nu} \tilde V(r,\hat \nu)\right|_{\hat \nu=\hat\nu_0}=0$ leads to the solution $\hat\nu_0=K(r)$. The original Hamiltonian~(\ref{inco}) can then be recovered since $\tilde V(r,\hat\nu_0)=V(r)$. The essence of the AFM is then to replace this function $\hat\nu_0$ by the constant $\nu_0$. In \cite{af}, it is shown that $\nu_0\approx \left\langle \Psi_A(\nu_0)\right|\hat\nu_0\left|\Psi_A(\nu_0)\right\rangle$. The AFM can consequently be regarded as a ``mean field approximation" with respect to a particular auxiliary field which is introduced to simplify the calculations. The main technical problem of the AFM is the determination of an analytical solution for~(\ref{rdef}) and (\ref{enmin}). Such a task has already been fulfilled for nonrelativistic Hamiltonians, either with central potentials of the form $a\, r^\lambda+b\, r^\eta$ \cite{af2} or with the exponential and Yukawa potentials \cite{af3}. 

An estimation of the error on an exact eigenvalue $E$ can be obtained. In \cite{af}, it is shown that
\begin{equation}
\label{errana2}
E(\nu_0)-E \gtrsim V(r_0)-\langle \Psi_A(\nu_0)| V(r) |\Psi_A(\nu_0)\rangle.
\end{equation}
The r.h.s.\ of this equation is the difference between the value of potential $V$ computed at the ``average point" $r_0$ and the average of this potential for the trial state. As we are mainly interested in the computation of eigenvalues, no evaluation of the quality of the approximate eigenstate $|\Psi_A(\nu_0)\rangle$ has been performed. This will be studied in a subsequent paper. 

An important result concerning the AFM has to be pointed out: This technique is equivalent to the envelope theory \cite{env0}, which is an other method to compute approximate analytical energy formulas \cite{afmenv}. As a consequence of this equivalence, the energy formulas which are found within both frameworks are identical. Since many results concerning semirelativistic Hamiltonians have already been obtained by resorting to envelope theory \cite{env,env2} one could wonder if it is still relevant to perform the present study. We claim that the answer is positive, mainly because in \cite{env,env2}, the energy formulas are all written under the form $E=\min_\nu[f(\nu)]$. The minimization stage~(\ref{enmin}), which is generally non trivial, is thus not explicitly performed with the consequence that no closed analytical formula is given; we achieve this stage in the present work in order to have completely explicit formulas. Moreover, we consider in the following two semirelativistic Hamiltonians that have not been studied up to now: the ones with square root and Yukawa potentials, both of interest in theoretical physics. A last argument is that we also propose an improvement of the energy formulas thanks to a comparison with the exact numerical data as it is done in \cite{af,af2,af3}. So, we are finally led to more accurate analytical formulas.        

\subsection{Useful results}\label{useres}

We have shown in \cite{af} that the eigenvalues of 
\begin{equation}
\label{eq:hlambda}
H_{\lambda}(a) = \frac{\bm{p}^{\, 2}}{\nu} + a\, \textrm{sgn}(\lambda) r^\lambda,
\end{equation}
where $\textrm{sgn}(\lambda) = \lambda/|\lambda|$ is the sign of $\lambda$ and where $a$ is a positive parameter, can be written, by using the AFM, under the form
\begin{equation}
\label{eq:eigenerplpot}
e_\lambda(a,\nu,N)=\frac{2+\lambda}{\lambda} \left|\frac{a \lambda}{2}\right|^\frac{2}{\lambda+2}
\left ( \frac{N^2}{\nu} \right )^\frac{\lambda}{\lambda+2}.
\end{equation}
$N$ depends on the radial $n$ and orbital $l$ quantum numbers. An important result, proved in \cite{af2}, is that the functional form of~(\ref{eq:eigenerplpot}) does not depend on the potential $P(r)$, provided that this potential is of power-law form. Only $N$ actually depends on the particular form of $P(r)$.  
With the choice $P(r)=r^2$, $N=2n+l+3/2$. Using the results from \cite{afmenv}, it is then possible to show that (\ref{eq:eigenerplpot}) gives an upper bound of the exact result. With the choice $P(r)=-1/r$, $N=n+l+1$. A lower bound is then yielded by (\ref{eq:eigenerplpot}). Since the form of $N$ is an artifact of the AFM, it can be modified to better fit the exact results. In particular, we have shown that, for values of $\lambda \in [-1,2]$, a good form for $N$ is given by \cite{af}
\begin{equation}
\label{eq:grandNl}
N_\lambda = b(\lambda)\, n + l + c(\lambda),
\end{equation}
with 
\begin{eqnarray}
\label{bc3}
\fl
b(\lambda)= \frac{(4\omega-18) \lambda+(18-2\omega)}
{(3\omega-15) \lambda+(21-3\omega)}, \quad
c(\lambda)= \frac{(7\omega-36) \lambda+(36-5\omega)}{(6\omega-32) \lambda+(40-6\omega)},
\end{eqnarray}
and $\omega=\sqrt{3}\pi $. Such expressions for the functions $b(\lambda)$ and $c(\lambda)$ lead to energy formulas as precise as $10^{-3}$ for the most interesting (the lowest) values of the quantum numbers $n$ and $l$ \cite{af}. But the variational character of (\ref{eq:eigenerplpot}) cannot then be guaranteed. Notice that the energy formula~(\ref{eq:eigenerplpot}) together with~(\ref{eq:grandNl}) and (\ref{bc3}) gives the exact result for the harmonic ($N_2=2n+l+3/2$) and Coulomb ($N_{-1}=n+l+1$) potentials and also reproduces the exact asymptotic behavior of the linear case ($N_1=\frac{\pi}{\sqrt 3}\, n+l+\frac{\pi\sqrt3}{4}$) \cite{af}.

We also found in \cite{af2} an accurate analytical energy formula for the Hamiltonian
\begin{equation}
\label{eq:funpot}
H=\frac{\bm p^2}{\nu}+a r - \frac{b}{r}.
\end{equation}
This linear-plus-Coulomb potential, also called the funnel or Cornell potential, is of great importance in hadronic  physics. It corresponds to a linear confinement associated with a
short range Coulomb contribution coming from one-gluon exchange processes \cite{lucha}, and has been shown to emerge as the effective energy between a static quark-antiquark pair in lattice QCD \cite{bali}. Defining 
\begin{equation}
\label{eq:Yfun}
Y = 3 N^2 \sqrt{ \frac{3a}{\nu^2 b^3}},
\end{equation}
one can express the energy spectrum of (\ref{eq:funpot}) as \cite{af2}
\begin{equation}
\label{eq:Efun}
e_{\textrm{f}} 
= \sqrt{3 a b} \left( \frac{Y}{F^2_+(Y)} - \frac{2}{F_+(Y)} \right),
\end{equation}
with $F_+(Y)$ given by (\ref{eq:rootcubeq}). With the choice $N=N_2$ ($N_{-1}$), (\ref{eq:Efun}) gives upper (lower) bounds. A better accuracy can be achieved with the choice $N=b(\beta)+l+c(\beta)$, where $\beta^4=\nu^2 b^3/(27 a)$.
The explicit form of $b(\beta)$ and $c(\beta)$ can be found in \cite{af2}. But again no information on the variational character can then be obtained.

As we have analytical and accurate energy formulas for any power-law potential and for the funnel potential at our disposal, we will assume in this work that the choices $P_{\lambda}(r)=\textrm{sgn}(\lambda)\, r^\lambda$ or $P_{\textrm{f}}(r)=ar-b/r$ are good starting points to apply the AFM.

\subsection{Alternative method}\label{altAFM}

There exists another way to introduce auxiliary fields that has already been presented in \cite{afhyb}; we recall and extend it in this section. This alternative method is formally simpler but less general than the AFM presented in section~\ref{geneAFM}. Nevertheless, it can lead to analytical results in many relevant cases and will appear to be of crucial importance to find analytical eigenvalues of semirelativistic Hamiltonians. 

Let us suppose that the Hamiltonian of our problem is the sum of $Q$ different contributions
\begin{equation}\label{hamA1}
	H_0=\sum^Q_{i=1}K_i,
\end{equation}
where $K_i$ are some operators. The eigenvalues and eigenstates of $H_0$ are denoted $E_0$ and $\left|\psi_0\right\rangle$ respectively. We can introduce $k$ auxiliary fields ($k \le Q$), denoted as $\phi_i$, to obtain a new Hamiltonian $H_k$ as follows
\begin{equation}\label{hamA2}
	H_k(\phi_m)=\sum^k_{i=1}\left(\alpha_i\frac{K^2_i}{\phi_i}+\beta_i\phi_i\right)+\sum^Q_{j=k+1}K_j,
\end{equation}
with real numbers $\alpha_i$, $\beta_i$ such that $\alpha_i\beta_i=1/4$.
Since the idea is to replace an operator by its square, this method can be expected relevant only for definite positive operators. In this case, the quantities $\alpha_i$, $\beta_i$, and $\phi_i$ must also be definite positive. 
The auxiliary fields $\phi_m$ should be rigorously eliminated as operators thanks to the constraint $\left.\delta_{\phi_i}H_k(\phi_m)\right|_{\phi_i=\hat\phi_i}=0$. One then obtains 
\begin{equation}\label{meanf}
\hat\phi_i=\sqrt{\alpha_i/\beta_i}\, K_i,	
\end{equation}
and $H_k(\hat\phi_j)=H_0$. Consequently, Hamiltonians~(\ref{hamA1}) and (\ref{hamA2}) are equivalent up to a proper elimination of the auxiliary fields as operators. 

We now consider that $\phi_i$ are variational parameters in order to simplify the calculations and denote $E_k(\phi_m)$ and $\left|\psi_k(\phi_m)\right\rangle$ the eigenvalues and eigenstates of $H_k(\phi_m)$. Then the set of coupled  equations $\left.\partial_{\phi_i}E_k(\phi_m)\right|_{\phi_m=\phi_{m;0}}=0$ have to be solved and the final eigenenergies and eigenstates read $E_{k;0}=E_k(\phi_{m;0})$ and $\left|\psi_{k;0}\right\rangle=\left|\psi_k(\phi_{m;0})\right\rangle$ respectively. An example where this procedure can be useful is the linear potential: No analytical solution can be found in general for the Schr\"{o}dinger equation with $V(r)=a\, r$, but the replacement of $V(r)$ by $\alpha V^2(r)/\phi+\beta\phi$ [see~(\ref{hamA2})] only involves a harmonic oscillator and an analytical solution is then possible to find. Another application which will be used below is the elimination of the square root operator in (\ref{ham0}) \cite{sema04}.

The Hellmann-Feynman theorem \cite{feyn} states that 
\begin{equation}
\partial_{\phi_i}E_k(\phi_m) = \left\langle\psi_k(\phi_m)\right| \partial_{\phi_i} H_k(\phi_m) \left|\psi_k(\phi_m)\right\rangle.
\end{equation}
The computation of this equation for the fields $\phi_{i;0}$ gives
\begin{equation}
\label{defphi2}
\phi_{i;0}^2=\frac{\alpha_i}{\beta_i} \left\langle\psi_{k;0}\right| K_i^2 \left|\psi_{k;0}\right\rangle
= \left\langle\psi_{k;0}\right| \hat \phi_i^2 \left|\psi_{k;0}\right\rangle.
\end{equation}
It is then possible to obtain an analytical form for the approximate energy
\begin{eqnarray}
\label{Ek0}
E_{k;0}&=& 2\sum^k_{i=1} \beta_i \phi_{i;0} + \sum^Q_{j=k+1}\left\langle\psi_{k;0}\right| K_j \left|\psi_{k;0}\right\rangle \nonumber \\
&=& \sum^k_{i=1} \sqrt{\left\langle\psi_{k;0}\right| K_i^2 \left|\psi_{k;0}\right\rangle} + \sum^Q_{j=k+1}\left\langle\psi_{k;0}\right| K_j \left|\psi_{k;0}\right\rangle .
\end{eqnarray}
The result is then independent of $\alpha_i$ and $\beta_i$, which can be chosen at best convenience. 
At first sight, this equation could be seen as a triviality, but the problem is to determine $\left|\psi_{k;0}\right\rangle$. This is precisely the purpose of the AFM. In the rest of this section, we will denote $\left\langle\psi_{k;0}\right| O \left|\psi_{k;0}\right\rangle$ by $\left\langle O \right\rangle$. Let us remark that (\ref{defphi2}) confirms the mean-field interpretation of the AFM.

This eigenenergy $E_{k;0}$ can be rewritten as 
\begin{equation}
E_{k;0}=\sum^k_{i=1} \left( \sqrt{\left\langle K_i^2 \right\rangle}-\left\langle K_i \right\rangle \right)+\left\langle  H_0 \right\rangle.
\end{equation}
Using the trivial inequality 
\begin{equation}
\left( K_i-\frac{1}{2 \alpha_i} \phi_i \right)^2 \ge 0,
\end{equation}
together with $\alpha_i \beta_i =1/4$, it is easy to show that
\begin{equation}
K_i \le \alpha_i\frac{K^2_i}{\phi_i}+\beta_i\phi_i.
\end{equation}
So, we have $H_k (\phi_m) \ge H$. From \cite{macd33} and \cite{srcoul3}, we can conclude that $\left\langle H_0 \right\rangle\geq E_0$ for any state. It follows that
\begin{equation}
\label{Ek02}
E_{k;0} \ge E_0 + \delta \quad \textrm{with} \quad \delta=\sum^k_{i=1} \left( \sqrt{\left\langle K_i^2 \right\rangle}-\left\langle K_i\ \right\rangle \right).
\end{equation}
It is easy to see that 
\begin{equation}
0 \le 
\left\langle \left( K_i - \sqrt{\left\langle K_i^2 \right\rangle} \right)^2 \right\rangle=
2 \sqrt{\left\langle K_i^2 \right\rangle} \left( \sqrt{\left\langle K_i^2 \right\rangle}-\left\langle K_i \right\rangle \right),
\end{equation}
which shows that $\delta \ge 0$.
This implies that the energy spectrum obtained by introducing auxiliary fields, as described by the above procedure, is an upper bound of the exact energy spectrum. A lower bound on the difference $E_{k;0}-E_0$ is then given by $\delta$. 

\section{Semirelativistic Hamiltonians}\label{srhamproc}

The rest of this work is devoted to find analytical approximate energy formulas for semirelativistic Hamiltonians of the form
\begin{equation}\label{ssh}
	H=\sigma\, \sqrt{\bm p^2+m^2}+V(r),
\end{equation}
where the potential is assumed to be central, and where the real parameter $\sigma$ can take all positive values. Actually, $\sigma = 1$ or 2 are used for most practical problems. If we deal with a one-body problem in a central potential or if we consider a two-particle system with a particle of mass $m$ and a particle of mass $M \gg m$ (so that the dynamic of the massive body can be forgotten), we must take $\sigma=1$. On the other hand, in the two-body case, we restrict our study to systems where both constituents have the same mass $m$ and, in this case, $\sigma=2$. Comments about the unequal mass case are given in \ref{uneqma}. Although these two situations are the most usual, we find convenient to let $\sigma$ as a free real parameter; some examples of such a useful prescription will be given later. The relativistic nature of this last Hamiltonian comes from the fact that the kinetic energy has no longer a nonrelativistic form, \textit{i.e.} $\bm p^2/2m$, but a relativistic one. The operator $\sqrt{\bm p^2+m^2}$ is actually the relativistic kinetic energy of a free spinless particle of mass $m$. The neglect of spin effects is one of the reasons why we use the word ``semirelativistic" to characterize Hamiltonian~(\ref{ssh}). In the literature, such a Hamiltonian is usually referred to as a spinless Salpeter Hamiltonian. 

\subsection{Auxiliary field formalism}

Because of the square root appearing in~(\ref{ssh}), finding exact analytical eigenvalues seems to be hopeless, excepted in very particular cases (see section~\ref{srho}). However it is possible to resort to the alternative AFM presented in section~\ref{altAFM} to rewrite the spinless Salpeter Hamiltonian under an apparently nonrelativistic form. Let us set $K_1=\sigma\, \sqrt{\bm p^2+m^2}$ and $K_2=V(r)$ in~(\ref{hamA1}). Then, we can introduce one auxiliary field, denoted as $\phi_1=\nu$, with $\alpha_1=1/\sigma^2$ and $\beta_1=\sigma^2/4$, and rewrite (\ref{hamA2}) in this particular case as 
\begin{equation}\label{ssh2}
	H(\nu)=\frac{m^2}{\nu}+\frac{\sigma^2}{4}\nu+\frac{\bm p^2}{\nu}+V(r).
\end{equation}

The auxiliary field $\nu$ can be eliminated as an operator thanks to the constraint $\left.\delta_\nu H(\nu)\right|_{\nu=\hat\nu}=0$, leading to 
\begin{equation}\label{nuhat}
\hat\nu=\frac{2}{\sigma}\sqrt{\bm p^2+m^2}.
\end{equation}
Obviously, one has then $H(\hat\nu)=H$, that is the equivalence between Hamiltonians~(\ref{ssh}) and (\ref{ssh2}). But, if $\nu$ is seen as a real number, (\ref{ssh2}) is simply a nonrelativistic Hamiltonian with kinetic term $\bm p^2/\nu$, for which analytical solutions can be hoped. To our knowledge, this technique has firstly been used in \cite{lucha2} but without explicit reference to the AFM. If $e(\nu)$ are the eigenvalues of $h(\nu)=\bm p^2/\nu+V(r)$, one can write the eigenvalues of $H(\nu)$ as
\begin{equation}\label{enu1}
	E(\nu)=\frac{m^2}{\nu}+\frac{\sigma^2}{4}\nu+e(\nu),
\end{equation}
  and the final spectrum is given by $E(\nu_0)$, where $\nu_0$ is such that 
\begin{equation}\label{enu2}
\left. \partial_\nu E(\nu)\right|_{\nu=\nu_0}=0\Rightarrow \frac{\sigma^2}{4}+e'(\nu_0)=\frac{m^2}{\nu^2_0}.
\end{equation}
The prime denotes a derivation with respect to $\nu$. Provided that $e(\nu)$ is analytically known, the only difficulty in this procedure is the analytical resolution of this last equation, especially when $m\neq0$. Reporting the value of $\nu_0$ given by (\ref{enu2}) in the expression of the energy (\ref{enu1}) allows to write
\begin{equation}\label{enu4}
E(\nu_0) = \frac{\sigma^2}{2}\nu_0+\left.\left(\nu\, e(\nu)\right)'\right|_{\nu=\nu_0}.
\end{equation}

With the notations used in this section, $E$ is an exact solution of the Hamiltonian $H$ (\ref{ssh}) and $E(\nu)$ is an exact solution of the Hamiltonian $H(\nu)$ (\ref{ssh2}). From the results of section~\ref{altAFM}, we know that $E(\nu) \ge E$ for any state, and that $E(\nu_0) \ge E$, since it is also true for the particular value $\nu_0$ of the auxiliary field. Let $E^*(\nu)$ be a solution (often approximate) of the Hamiltonian~(\ref{ssh2}) obtained by the AFM. If $E^*(\nu) \ge E(\nu)$, then we have $E^*(\nu_0) \ge E$, and an upper bound is available for each eigenstate of the Hamiltonian~(\ref{ssh}). If $E^*(\nu) \le E(\nu)$ or if it is not possible to locate $E^*(\nu)$ with respect to $E(\nu)$, no information can be obtained for the variational character of $E^*(\nu_0)$. The value of $\nu_0$ is given by
\begin{equation}\label{enu3}
\nu_0=\frac{2}{\sigma}\sqrt{\left\langle \bm p^2+m^2 \right\rangle},
\end{equation}
where the mean value is computed with an eigenstate of $H(\nu_0)$.
An estimation of the difference between the exact and approximate eigenenergies is given by~(\ref{Ek02}), which reads in this case
\begin{equation}
\delta=\sigma \sqrt{\left\langle \bm p^2+m^2 \right\rangle}- \sigma \left\langle \sqrt{\bm p^2+m^2} \right\rangle.
\end{equation}

At the limit of low mass, (\ref{ssh}) can be written 
\begin{equation}\label{sshlowm}
H\approx H^{\textrm{ur}}+\frac{\sigma\, m^2}{2 \sqrt{\bm p^2}} \quad \textrm{with} \quad
H^{\textrm{ur}}=\sigma\, \sqrt{\bm p^2}+V(r).
\end{equation}
The contribution $\Delta(m)$ of the mass $m$ to an eigenenergy of the ultrarelativistic Hamiltonian $H^{\textrm{ur}}$ appears as a small contribution that can be computed as a perturbation. The mean value being taken with a eigenfunction of $H^{\textrm{ur}}$, we have 
\begin{equation}\label{mcontur}
\Delta(m)=\left\langle \frac{\sigma\, m^2}{2 \sqrt{\bm p^2}}\right\rangle \approx\frac{\sigma\, m^2}{2 \sqrt{\left\langle\bm p^2\right\rangle}} = \frac{m^2}{\bar \nu_0},
\end{equation}
where $\bar \nu_0$ is the auxiliary field obtained with the AFM applied to the Hamiltonian $H^{\textrm{ur}}$. The validity of this approximation will be checked in section~\ref{subsec:lowmass}.

\subsection{Scaling laws}

Let us consider the spinless Salpeter equation
\begin{equation}
\label{sl1}
\left(\sigma\, \sqrt{\bm p^2+m^2}+G\, V(a\, \bm r)-E(m,G,a,\sigma)\right)\, \Psi(\bm r\,)=0,
\end{equation}
where $G$ and $a$ are two real numbers. Defining $\bm \rho=a\bm r$, $\bm\pi=\bm p/a$, one has
\begin{equation}
\label{sl2}
\left(\sqrt{\bm \pi^2+(m/a)^2}+\frac{G}{a\sigma}\, V(\bm \rho)-\frac{1}{a\sigma}E(m,G,a,\sigma)\right)\, \Psi(\bm \rho/a)=0.
\end{equation}
This is just the spinless Salpeter equation 
\begin{equation}
\label{sl3}
\left(\sqrt{\bm \pi^2+m'^2}+G'\, V(\bm \rho)-E'(m',G',1,1)\right)\, \varphi(\bm \rho)=0,
\end{equation}
with 
\begin{equation}
	m'=\frac{m}{a},\quad G'=\frac{G}{a\sigma},\quad E'(m',G',1,1)=\frac{E(m,G,a,\sigma)}{a\sigma}.
\end{equation}
We are thus led to the following scaling law for the energy spectrum
\begin{equation}\label{scal}
	E(m,G,a,\sigma)=a\, \sigma\ E\left(\frac{m}{a},\frac{G}{a\sigma},1,1\right).
\end{equation}
It means that both one parameter of the potential and the number of constituents in the system can always be set equal to 1 in order to simplify the computations without loss of generality. The general energy formula will then be recovered thanks to~(\ref{scal}).

It is worth mentioning that in the particular case of a power-law potential, \textit{i.e.} $V(r)=G\, \textrm{sgn}(\lambda)
r^\lambda$, the same computations as above can be performed and lead to a more constraining scaling law that reads
\begin{equation}
E_\lambda(m,G,\sigma)=(\sigma^\lambda\, G)^{\frac{1}{\lambda+1}}\, E_\lambda\left(\chi=\left(\frac{\sigma}{G}\right)^{\frac{1}{\lambda+1}}\, m,1,1\right).	
\end{equation}
This means that the whole power-law problem can be studied by only considering the eigenvalue $E_\lambda(\chi,1,1)$ associated to the two-parameter Hamiltonian
\begin{equation}
	{\cal H}_\lambda(\chi)=\sqrt{\bm \pi^2+\chi^2}+\textrm{sgn}(\lambda) \rho^\lambda.
\end{equation}

\section{Power-law potentials}\label{plpot}
\subsection{General case}
We apply in this section the AFM to find approximate analytical energy formulas for the spinless Salpeter Hamiltonian (\ref{ssh}) with the potential $V_\lambda(r)=a\, \textrm{sgn}(\lambda)
r^\lambda$ where $a$ is positive. We consider that $\lambda>-2$, as needed in the nonrelativistic case to have bound states. 

First, we can get rid of the square root appearing in the kinetic energy by introducing one auxiliary field $\nu$ as it is done in~(\ref{ssh2}). We obtain
\begin{equation}
	H_\lambda(\nu)=\frac{m^2}{\nu}+\frac{\sigma^2}{4}\nu+\frac{\bm p^2}{\nu}+a\, \textrm{sgn}(\lambda)
r^\lambda.
\end{equation}
If $\nu$ is assumed to be a real number, this last Hamiltonian is simply a nonrelativistic one with a power-law potential, for which analytical energy formulas were obtained in \cite{af} and recalled in section~\ref{useres}. Using the result~(\ref{eq:eigenerplpot}) one is led to
\begin{equation}\label{epl1}
\fl
E_\lambda(\nu)=\frac{m^2}{\nu}+\frac{\sigma^2}{4}\nu+ \frac{\lambda+2}{\lambda}\, A_\lambda\, \nu^{-\frac{\lambda}{\lambda+2}}\quad  {\rm with} \quad 
A_\lambda=\left|\frac{a\, \lambda}{2}\right|^{\frac{2}{\lambda+2}}\, N^{\frac{2\lambda}{\lambda+2}}.
\end{equation}
At this stage, setting $N=N_\lambda$ seems a good choice (see (\ref{eq:grandNl})), but it is not necessarily the best one in the present case. Indeed, $N_\lambda$ has been fitted on the energy levels of the exact nonrelativistic spectrum, while we deal here with a relativistic kinematics. Thus, we can expect that the best expression for $N$ is different from $N_\lambda$. Inspired by (\ref{eq:grandNl}), one could try $N=b(m,\lambda) n+l+c(m,\lambda)$ (see section~\ref{increase}).

The auxiliary field $\nu$ can now be eliminated by minimizing~(\ref{epl1}) with respect to $\nu$, the minimum being obtained for the value $\nu=\nu_0$. Setting
\begin{equation}
	\nu_0=x_0^{\frac{\lambda+2}{2}},
\end{equation}
we obtain
\begin{equation}\label{epl3}
	\left.\partial_{\nu}E_\lambda(\nu)\right|_{\nu=\nu_0}=0\Rightarrow \frac{\sigma^2}{4}x_0^{\lambda+2}-A_\lambda\, x_0-m^2=0.
\end{equation}
Assuming that we know the solution $x_0$ from~(\ref{epl3}), two convenient forms can be found for the physical values $E_\lambda=E_\lambda(x_0^{\frac{\lambda+2}{2}})$ of the eigenenergies~(\ref{epl1})
\begin{equation}\label{enfin1}
	E_\lambda=\frac{\sigma}{\lambda}\frac{\lambda m^2+(\lambda+1)\, A_\lambda\, x_0}{\sqrt{m^2+A_\lambda\, x_0}},
\end{equation}
or
\begin{equation}\label{enfin2}
	E_\lambda=\frac{2}{\lambda}\left((\lambda+1)\frac{\sigma^2}{4}\, \nu_0-\frac{m^2}{\nu_0}\right).
\end{equation}

In order to obtain analytical energy formulas, one should be able to analytically solve (\ref{epl3}). Let us now examine the cases for which this is possible. First, we set $\lambda+2=p/q$ and $X_0=x_0^{1/q}$, where $p$ and $q$ are relatively prime. Then, (\ref{epl3}) becomes
\begin{equation}
	\frac{\sigma^2}{4}X_0^{p}-A_\lambda\, X_0^q-m^2=0.
\end{equation}
A polynomial possesses analytical roots if its order is less or equal to 4; therefore all the solvable potentials should verify the conditions $1 \leq p,\, q \leq 4$. An exhaustive research of all the solvable potentials
leads to the following values for the power $\lambda$:
\begin{equation}
\label{vallambr2}
\lambda = -2,\: -\frac{7}{4},\: -\frac{5}{3},\:- \frac{3}{2}, \: -\frac{4}{3},\: -\frac{5}{4},\: -1,\: -\frac{2}{3}, \: -\frac{1}{2},\: 0,\: 1,\: 2.
\end{equation}
Among these allowed values, three are of particular interest: $\lambda=-1$ and $2$ because the Coulomb problem and the harmonic oscillator play a central role in theoretical physics, and $\lambda=1$ since a linearly rising potential is generally considered to be a relevant approximation of the confining potential in QCD \cite{lucha}. These three cases are explicitly solved in the following sections.

Let us briefly discuss the existence of physical solutions, as function of the parameters. Starting from (\ref{epl3}) and (\ref{enfin1}), it is possible to prove the following properties:
\begin{itemize}
  \item For $\lambda > 0$, there exists always a physical solution with $E_\lambda > 0$;
  \item For $-1 \leq \lambda < 0$ there exists a physical solution with $E_\lambda > 0$ as long as the parameter $a$ is less than a critical value $a_c(\lambda)$ given by
\begin{equation}
\label{aclambda}
a_c(\lambda) = \sigma \left ( \frac{N}{\sqrt{|\lambda|}} \right )^{|\lambda|} \left ( \frac{m^2}{1+\lambda} \right )^\frac{1+\lambda}{2}.
\end{equation}
Let us remark that $a_c(\lambda)$ depends on $m$, except $a_c(-1)=\sigma\,N$.
  \item For $-2 \leq \lambda < -1$ either there is no solution for (\ref{epl3}), or there exist two solutions but in the latter case these solutions are not compatible with the nonrelativistic expressions when $m$ tends toward infinity. In both cases the corresponding solutions are not physical so that these types of potential must be discarded from our study.
\end{itemize}

\subsection{Harmonic oscillator}\label{harmo1}

If $\lambda=2$, (\ref{epl3}) becomes a quartic equation of the form
\begin{equation}
\label{quartic1}
	\frac{\sigma^2}{4}x^4_0-A_2\, x_0-m^2=0,\quad {\rm with}\quad A_2=\sqrt a\, N.
\end{equation}
Defining
\begin{equation}
x_0=\left(\frac{2A_2}{\sigma^2}\right)^{1/3}\, X\quad {\rm and}\quad Y_2=\frac{m^2}{3} \left(\frac{16\sigma}{a N^2}\right)^{2/3},
\end{equation}
(\ref{quartic1}) can be rewritten as
\begin{equation}\label{ho1}
	4X^4-8X-3Y_2=0,
\end{equation}
which is precisely of the form (\ref{eq:redcubeq2}). Following~(\ref{eq:rootquarteq}), the solution of (\ref{ho1}) is given by $X(Y_2)=G_{-}(Y_2)$, and, after a rearrangement of~(\ref{enfin1}), the energy spectrum reads (using (\ref{ho1}) under one of the equivalent forms)
\begin{eqnarray}
\label{E2_3f}
	E_2 &= &\sigma\, m \sqrt{\frac{3}{Y_2}}  \frac{Y_2+4G_-(Y_2)}{\sqrt{8G_-(Y_2)+3Y_2}} \\
	& = & \sigma\, m \sqrt{\frac{3}{Y_2}} \left( \frac{2}{G_-(Y_2)} + \frac{Y_2}{2 G_-^2(Y_2)} \right)\\
	& = & \frac{2 \sigma\, m}{\sqrt{3 Y_2}} \left(G_-^2(Y_2) + \frac{1}{G_-(Y_2)}\right).
\end{eqnarray}
With $N=N_2$ these formulas yield an upper bound of the energy, since this setting gives the exact solution for the corresponding nonrelativistic Hamiltonian.

In the nonrelativistic limit ($Y\to \infty$), the last equations reduce to
\begin{equation}\label{hoylarge}
	E_2 \approx \sigma\, m + \sqrt{\frac{2 \sigma a}{m}} N,
\end{equation}
as expected for a nonrelativistic harmonic oscillator (see (\ref{eq:eigenerplpot}) for $\lambda=2$ and $\nu=2m/\sigma$). For large value of $m$, the choice $N=N_2$ is clearly optimal. For small values of $m$, another choice could give better results.

\subsection{Linear potential}\label{line1} 

The resolution of the case $\lambda=1$ is rather similar to the one of the harmonic oscillator. Defining
\begin{equation}
	x_0=2\left(\frac{A_1}{3\sigma^2}\right)^{1/2} X\quad{\rm and}\quad Y_1=\frac{3^{3/2}m^2\sigma}{4A_1^{3/2}},
\end{equation}
(\ref{epl3}) simply becomes a cubic equation of the form (\ref{eq:redcubeq1}), that is
\begin{equation}\label{lin1}
	X^3-3X-2Y_1=0.
\end{equation}
Notice that $A_1=\left(a N/2\right)^{2/3}$ so that
\begin{equation}
Y_1=\frac{3}{2} \frac{\sqrt{3} \sigma m^2}{a N}.
\end{equation}
Following~(\ref{eq:rootcubeq}), the solution of (\ref{lin1}) is given by $X(Y_1)=F_-(Y_1)$, and, after a rearrangement of~(\ref{enfin1}), the energy spectrum reads (using (\ref{lin1}) under one of the equivalent forms)
\begin{eqnarray}
\label{E1_3f}
	E_1 & = & \sigma\, m \sqrt{\frac{2}{Y_1}} \frac{Y_1+3F_-(Y_1)}{\sqrt{3F_-(Y_1)+2Y_1}} \\
	& = &\sigma\, m \sqrt{\frac{2}{Y_1 F_-(Y_1)}} \left( 3 + \frac{Y_1}{F_-(Y_1)} \right) \\
	& = & \frac{\sigma\, m }{\sqrt{2 Y_1 F_-(Y_1)}} \left( 3 + F_-^2(Y_1)\right). 
\end{eqnarray}

In the nonrelativistic limit ($Y\to \infty$), the last equations reduce to
\begin{equation}\label{hoylarge2}
	E_1 \approx \sigma\, m + \frac{3}{2} \left( \frac{\sigma a^2 N^2}{m} \right)^{1/3} ,
\end{equation}
as expected for a nonrelativistic Hamiltonian with a a linear potential (see (\ref{eq:eigenerplpot}) for $\lambda=1$ and $\nu=2m/\sigma$). For large value of $m$, the choice $N=N_1$ is a good one. For small values of $m$, another choice could give better results.

\subsection{The Coulomb problem}\label{coulsr}

Equation~(\ref{epl3}) is considerably simplified when $\lambda=-1$; the value of $x_0$ can directly be extracted in this case and reads
\begin{equation}
	x_0=\frac{4m^2}{\sigma^2-4A_{-1}},\quad  {\rm with}\quad A_{-1}=\left(\frac{a}{2N}\right)^2.
\end{equation}
The energy spectrum (\ref{enfin1}) is finally given by $E_{-1}=m \sqrt{\sigma^2-4A_{-1}}$, or
\begin{equation}\label{encoulsr}
	E_{-1}=\sigma m \sqrt{1-\frac{a^2}{\sigma^2N^2}}.
\end{equation}
It is obvious from this last equation that bound states exist only when $a < \sigma\, N$, 
in agreement with (\ref{aclambda}). The most stringent upper bound for $a$ is actually found for the ground state, that is $a<\sigma \left.N\right|_{n=l=0}$. With $N=N_{-1}$, (\ref{encoulsr}) yields an upper bound of the energy since this setting gives the exact solution for the corresponding nonrelativistic Hamiltonian.

As the coupling constant $a$ is dimensionless in this case, the only mass scale of the Hamiltonian is $m$. So the nonrelativistic limit cannot be obtained by setting $m\to\infty$, as usual. It is well known that the mean speed of the particle is independent of $m$ and proportional to $a$. So, the nonrelativistic limit is achieved for $a\to 0$. When $a\ll 1$, (\ref{encoulsr}) reduces to
\begin{equation}\label{encoulsrasmall}
	E_{-1}\approx \sigma m -\frac{a^2 m}{2\sigma N^2},
\end{equation}
which corresponds to the rest energy plus the Coulomb binding energy, as expected (see (\ref{eq:eigenerplpot}) for $\lambda=-1$ and $\nu=2m/\sigma$). For large values of $m$, the choice $N=N_{-1}$ is clearly optimal. For small values of $m$, another choice could give better results.

Several works have been devoted to the spinless Salpeter equation with a Coulomb potential \cite{srcoul1,srcoul2,srcoul3,srcoul4}, and it is interesting to compare our results with previously found ones. First of all, it has been shown in \cite{srcoul1} that the energy spectrum of the semirelativistic Coulomb problem is unbounded from below if 
\begin{equation}
	a> a_c = \frac{2\sigma}{\pi}.
\end{equation}
Moreover, a lower bound on the ground state energy $E_g$ is \cite{srcoul1}
\begin{equation}
	E_g\geq \sigma m\sqrt{1-\frac{a^2}{a_c^2}}.
\end{equation}
But, since we used the alternative AFM to get rid of the square root in the kinetic term, ~(\ref{encoulsr}) is an upper bound of the exact energy. This is confirmed by the results of \cite{srcoul3} in which formula~(\ref{encoulsr}) was already given.

A last result of interest is the analytical determination of the ground state energy performed in \cite{srcoul4}. It is found that 
\begin{equation}\label{ground}
	E_g |_{a=a_c} \lesssim \sigma\,m\times 0.4842564\dots
\end{equation}
Our formula~(\ref{encoulsr}) leads in this case to 
\begin{equation}
E_{-1}|_{a=a_c} = \sigma m\sqrt{1-\frac{4}{\pi^2\left.N\right|_{n=l=0}^2}} .
\end{equation}
This last expression is equal to 0.77$\, m$ if $N = N_{-1}$, but agrees with (\ref{ground}) if $\left.N\right|_{n=l=0}\approx 0.73$ is taken (see \ref{sec:rcoulp} for more information). 

\section{Ultrarelativistic limit}
\label{secur}

In the ultrarelativistic limit, that is to say $m=0$, the various elimination equations for the auxiliary field and the energy formulas become simpler. Before going further, we stress that the results presented in this section are only valid for $m=0$.
\subsection{Power-law potentials}
Equation~(\ref{epl3}) can be written in the ultrarelativistic limit as
\begin{equation}
x_0 \, \left(\frac{\sigma^2}{4}\, x_0^{\lambda+1}-A_\lambda\right)=0.	
\end{equation}
The nontrivial solution of this equation is 
\begin{equation}
	x_0=\left(\frac{\sigma^2}{4}\right)^{-\frac{1}{\lambda+1}}\, A_\lambda^{\frac{1}{\lambda+1}},
\end{equation}
and the final energy spectrum is given by
\begin{equation}\label{ezerom}	
E^{{\rm ur}}_\lambda(a,\sigma,N)=\frac{\lambda+1}{\lambda} \left|a\, \lambda\right|^{\frac{1}{\lambda+1}}\, \left(\sigma N\right)^{\frac{\lambda}{\lambda+1}}.
\end{equation}
Let us stress that, in this special case, an analytical expression is obtained whatever the power $\lambda$ of the potential, in contrast to the particular values (\ref{vallambr2}) resulting from the general case. One can check that (\ref{E2_3f}) and (\ref{E1_3f}) reduces to (\ref{ezerom}) when $m\to 0$.

For the particular case $\lambda=1$, this last formula becomes $E_1^{{\rm ur}}=2\sqrt{\sigma\, a\, N}$, and one has consequently $(E^{{\rm ur}}_1)^2\propto N$. Such a linear behavior between the squared eigenenergies and the quantum numbers of the different states (with $N=N_1$ for instance) is a well-known property of the spinless Salpeter Hamiltonian with a linear potential and massless constituents. Such a Hamiltonian is one of the simplest ways to describe a light meson in a potential approach of QCD, and it is experimentally checked that the squared masses of the light mesons mainly grow linearly with their angular momentum (Regge trajectories). See for example \cite{lucha} for a discussion of that point.  

The quantity $E^{{\rm ur}}_\lambda$ is physically a mass. It appears that $E^{{\rm ur}}_\lambda <0$ when $-1 < \lambda < 0$ and $E^{{\rm ur}}_{-1}=0$. No bound state of massless constituent particles can be found in these cases. Although (\ref{ezerom}) is positive for $-2<\lambda<-1$, this range of values for $\lambda$ has been proved to be unphysical.

By comparison of the nonrelativistic result (\ref{eq:eigenerplpot}) and the ultrarelativistic one (\ref{ezerom}), it can  be checked that 
\begin{equation}\label{dual}
	e_{2\lambda}(a^{\frac{1+2\lambda}{1+\lambda}},a^{\frac{1}{1+\lambda}}/\sigma,\sqrt{N})=E^{{\rm ur}}_\lambda(a,\sigma,N).
\end{equation}
There is a kind of duality between the nonrelativistic and ultrarelativistic energy formulas for power-law potentials, at least for the approximate relations given by the AFM.

\subsection{A special case: the Harmonic oscillator}\label{srho}

The energy spectrum of the Hamiltonian 
\begin{equation}\label{hamho}
  H_{{\rm ho}}=2\sqrt{\bm p^{\,2}}+a\, r^2
\end{equation}
can be analytically computed for $l=0$ and reads \cite{luchaho}
\begin{eqnarray}\label{srhowf}
E_{{\rm ho}}&=&-(4a)^{1/3}\alpha_n,
\end{eqnarray}
where $\alpha_n<0$ are the zeros of the regular Airy function ($n=0$, 1, \dots). They can be found for
example in \cite{Abra}, where the following WKB approximation of these zeros is also given 
\begin{equation}\label{dh1}
\alpha_n\approx-\left(\frac{3\pi}{2}\left(n+\frac{3}{4}\right)\right)^{2/3}\,.
\end{equation}
This approximation is precise up to $8\%$.
With the WKB expression (\ref{dh1}), an accurate approximation of the energy levels (\ref{srhowf}) reads 
\begin{equation}
	E_{{\rm ho}}=3\left(\sqrt a\,\left(\frac{\pi}{\sqrt 3}\, n+\frac{\pi\sqrt3}{4}\right)\right)^{2/3}.
\end{equation}

This last expression can be compared to the AFM result (\ref{ezerom}) for $\lambda=\sigma=2$, \textit{i.e.}
\begin{equation}	
E^{{\rm ur}}_2=3\left(\sqrt a\, N\right)^{2/3}.
\end{equation}
With (\ref{hamho}) as a starting point, a natural choice for $N$ is $N_2$, but 
it is clear that both $E_{{\rm ho}}$ and $E^{{\rm ur}}_2$ are identical provided that 
\begin{equation}\label{N2def}
	\left.N\right|_{l=0}=\frac{\pi}{\sqrt 3}\, n+\frac{\pi\sqrt3}{4}. 
\end{equation}
These values for the coefficient of $n$ and the zero point energy are exactly those that optimize the AFM energy formula in the case of a nonrelativistic kinetic energy with linear potential (see~(\ref{bc3})). The explanation of this fact is that the Fourier transform of Hamiltonian (\ref{hamho}) is a nonrelativistic Hamiltonian with linear potential. Consequently, it is not surprising that we find the expression (\ref{N2def}) for $\left.N\right|_{l=0}$. This is one more indication that $N$ should be different in the semirelativistic and nonrelativistic cases.

\section{Square root potential}\label{sqrtpot}

The square root potential $V(r)=\sqrt{a^2 r^2+b^2}$ plays an important role in the theoretical description of hybrid mesons \cite{hyb}. In the flux tube picture indeed, an excitation of the gluonic field gives rise to an excited flux tube potential of such a square root form. That is why we find interesting to show that analytical energy formulas for this potential can be easily obtained with the AFM. 

\subsection{Nonrelativistic kinematics}

The Fourier transform of Hamiltonian $H=\frac{\bm p^2}{2 m}+\sqrt{a^2 r^2+b^2}$, denoted as ${\cal F}(H)$, reads
\begin{equation}
	{\cal F}(H)=\sigma \sqrt{\bm p^2+M^2}+\kappa\,r^2, 
\end{equation}
with 
\begin{equation}
	\sigma=\left( \frac{4 a}{m^2} \right)^{1/3},\quad M=\frac{b}{\sigma},\quad \kappa = \sigma\frac{m a}{8}.
\end{equation}
This last Hamiltonian is nothing else than a spinless Salpeter one~(\ref{ssh}) with a harmonic potential.
Provided that the proper substitution rules are taken into account, the energy spectrum in this case is the same than the one computed in section~\ref{harmo1}. We have thus 
\begin{eqnarray}
	E^{{\rm nr}}_{{\rm sq}}&= &b \sqrt{\frac{3}{Y_2}}  \frac{Y_2+4G_-(Y_2)}{\sqrt{8G_-(Y_2)+3Y_2}} \\
	& = & b \sqrt{\frac{3}{Y_2}} \left( \frac{2}{G_-(Y_2)} + \frac{Y_2}{2 G_-^2(Y_2)} \right)\\
	& = & \frac{2 b}{\sqrt{3 Y_2}} \left(G_-^2(Y_2) + \frac{1}{G_-(Y_2)}\right),
\end{eqnarray}
with
\begin{equation}
Y_2=\frac{b^2}{3}\left(\frac{32 \,m}{a^2N^2}\right)^{2/3}.
\end{equation}
This potential has been studied in more detail in \cite{hybri,afsqrt}, and we recovered here the results presented in that works.
It can be checked that when $b\to 0$, these last equations reduce to the energy spectrum of a nonrelativistic Hamiltonian with linear potential, that is (\ref{eq:eigenerplpot}) with $\lambda=1$. 

\subsection{Ultrarelativistic limit}
The Fourier transform of $H=\sigma\sqrt{\bm p^2}+\sqrt{a^2 r^2+b^2}$ reads
\begin{equation}
	{\cal F}(H)=\sigma \sqrt{\bm p^2+M^2}+a\, r,
\end{equation}
with $M=b/\sigma$.
This last Hamiltonian appears to be a spinless Salpeter one~(\ref{ssh}) with a linear potential.
Provided that the proper substitution rules are taken into account, the energy spectrum in this case is the same than the one computed in section~\ref{line1}. We have thus 
\begin{eqnarray}
	E^{{\rm ur}}_{{\rm sq}} & = & b \sqrt{\frac{2}{Y_1}} \frac{Y_1+3F_-(Y_1)}{\sqrt{3F_-(Y_1)+2Y_1}} \\
	& = &b \sqrt{\frac{2}{Y_1 F_-(Y_1)}} \left( 3 + \frac{Y_1}{F_-(Y_1)} \right) \\
	& = & \frac{b}{\sqrt{2 Y_1 F_-(Y_1)}} \left( 3 + F_-^2(Y_1)\right), 
\end{eqnarray}
with
\begin{equation}
Y_1=\frac{3^{3/2}\, b^2}{2a\, \sigma N}.
\end{equation}
When $b\to 0$, these last equations logically reduce to (\ref{ezerom}) with $\lambda=1$. 

\section{Funnel potential}\label{funnelpot}

As we stressed in section~\ref{useres}, the funnel potential $V(r)=ar-b/r$, with $a$ and $b$ both positive, is of crucial importance in hadronic physics. It is thus an interesting potential to be studied with the AFM.
The energy spectrum of~(\ref{ssh2}) with the funnel potential is analytically known thanks to (\ref{eq:Efun}), and is given by
\begin{equation}
\label{eq:eigenfun}
E_{{\rm f}}(Y)=\frac{m^2}{U}\, Y+\frac{\sigma^2U}{4Y}+	\sqrt{3 a b} \left( \frac{Y}{F^2_+(Y)} - \frac{2}{F_+(Y)} \right),
\end{equation}
with
\begin{equation}
\label{YUnu}
Y = \frac{U}{\nu} \quad {\rm and}\quad  U=3 N^2 \sqrt{ \frac{3a}{b^3}}.
\end{equation}
Note that an alternative expression, often more convenient, for (\ref{eq:eigenfun}) is obtained thanks to the following relation
\begin{equation}
\label{eq:altexpressE}
 \frac{Y}{F^2_+(Y)} - \frac{2}{F_+(Y)}= \frac{1}{2} \left(F_+(Y) - \frac{1}{F_+(Y)} \right).
\end{equation}
It is more useful to express the energy as a function of $Y$ rather than as a function of $\nu$, since the analytical minimization with respect to $Y$ looks formally simpler than the same task with $\nu$.
After some algebra, the condition $\left. \partial_Y E(Y)\right|_{Y=Y_0}=0$ can be rewritten as 
\begin{equation}
\label{eq:detY0fun}
	\frac{\sqrt{3ab}}{3F^2_+(Y_0)}+\frac{m^2}{U}-\frac{\sigma^2 U}{4Y^2_0}=0.
\end{equation}
The property $F^3_+(Y)+3F_+(Y)-2Y=0$ has to be used.

\subsection{Massless particles}
\label{subsec:masslesspart}

In the ultrarelativistic limit ($m=0$), (\ref{eq:detY0fun}) becomes
\begin{equation}
\label{eq:defD}
	\frac{F_+(Y_0)}{Y_0}=\frac{2b}{3\sigma N}=D,
	\end{equation}
	whose solution is
\begin{equation}\label{eq:Y0D}
	Y_0=\sqrt{\frac{2-3D}{D^3}}.
\end{equation}
To obtain this result, the definition relation of $F_+(Y)$ is used with the special value $Y = Y_0$.
It is worth mentioning that one has the constraint $2-3D>0$ in order for $Y_0$ to be defined, or equivalently 
$b < \sigma N$.
This constraint is formally the same as for the pure Coulomb potential and states that $b$ cannot be arbitrarily large without forbidding bound states. This feature of the ultrarelativistic funnel case has numerically been observed in \cite{boul}.

Finally, the energy formula reads
\begin{equation}\label{funnelur}
	E^{{\rm ur}}_{{\rm f}}=	E_{{\rm f}}(Y_0)=2 \sqrt{a \left( \sigma N\ - b \right)}.
\end{equation}
It is astonishing that such a complicated potential used in conjunction with a semirelativistic kinetic energy admits analytical approximate eigenenergies of such simple form. For $b=0$, (\ref{funnelur}) reduces to $2\sqrt{a\sigma N}$, that is the expected expression from (\ref{ezerom}) in the case $\lambda=1$. For $a=0$, $E^{{\rm ur}} \to 0$ as shown in section~\ref{secur}.
If the prescription $N=N_2$ is chosen, then (\ref{funnelur}) is an upper bound of the exact result. As already mentioned, other choice for $N$ could improve the accuracy of the formula but without no guarantee about the variational character of the result.

\subsection{General case}
\label{subsec:gencas}

At first sight, solving (\ref{eq:detY0fun}), may seem hopeless. However, one can indeed obtain an analytical solution. The calculations are somewhat complicated, and we report here only the main steps and useful results. 

The first step consists in multiplying this equation by $F_+(Y_0)^3 = 2Y_0 - 3F_+(Y_0)$ (this is the result (\ref{eq:rootcubeq}) applied to the value $Y=Y_0$) in order to get an expression linear in the $F$ function. One is able to obtain the following relationship
\begin{equation}
\label{eq:expFY0}
F_+(Y_0)=\frac{2 Y_0 (1 - u^2 Y_0^2)}{c Y_0^2 + 3},
\end{equation}
where we introduced intermediate quantities
\begin{eqnarray}
\label{eq:defqtuc}
u & = & \frac{2m}{\sigma U} = \frac{2m}{3 \sigma N^2} \sqrt{\frac{b^3}{3a}}, \\
c & = & D^2 - 3 u^2 = \frac{4 b^2 A}{9 a \sigma^2 N^4}, \\
A & = & a N^2 - b m^2,
\end{eqnarray}
and where $D$ has already be defined in (\ref{eq:defD}).

Instead of using the variable $Y_0$, it is more convenient to work with the new variable $Z_0$ defined as
\begin{equation}
\label{eq:defZ0}
Z_0 = c Y_0^2,
\end{equation}
so that (\ref{eq:expFY0}) can be recast into the simpler form
\begin{equation}
\label{eq:newdefFY0}
F_+(Y_0)=\frac{2 Y_0 (1 - d Z_0)}{Z_0 + 3},
\end{equation}
with
\begin{equation}
\label{eq:defpetd}
d = \frac{u^2}{c} = \frac{b m^2}{3 A}.
\end{equation}
Let us remark the interesting relation $c + 3 c d = D^2$ which will be used in the following.

Inserting (\ref{eq:newdefFY0}) into the equation defining the $F_+$ function (\ref{eq:rootcubeq}), we get rid of the $F_+$ function and we remain with a polynomial in $Z_0$. Explicitly, we end up with the equation in $Z_0$ to be solved
\begin{equation}
\label{eq:eqgivZ0}
4 d^3 Z_0^3 + (D^2 - 12 d^2) Z_0^2 + 6(D^2+2d) Z_0 + 9 D^2 - 4 = 0.
\end{equation}
Since the polynomial is of third order, its root $Z_0$ can be expressed analytically. The same property holds for $Y_0$ through (\ref{eq:defZ0}), for $F_+(Y_0)$ through (\ref{eq:newdefFY0}) and lastly for the searched energy $E$ through (\ref{eq:eigenfun}) (the use of the alternative form (\ref{eq:altexpressE}) is interesting).

Although we have obtained the analytical expression for this general case, the resulting form is quite complicated and not reported in this paper, but it can be obtained on request. We prefer to give the simpler case valid for low masses.

\subsection{Low mass expansion}
\label{subsec:lowmass}

For $m=0$, it is easy to show that $u = d = 0$, $c=D^2$ so that the solution of (\ref{eq:eqgivZ0}) is $Z_0 = \bar{Z} = (2 - 3D)/D$ leading to $F_+(Y_0)= \bar{F} = \sqrt{\bar{Z}} = D Y_0 = D \bar{Y}$, and $E$ given by (\ref{funnelur}).
Let us define $m_0 = N \sqrt{a/b}$. By low mass, we mean the condition $m \ll m_0$, or $\epsilon = (m/m_0)^2 \ll 1$. In such a case, $A = a N^2 (1 - \epsilon)$ so that $c = D^2 (1 -\epsilon)$ and $d= \epsilon (1-\epsilon)^{-1}/3$. Keeping only the terms of first order in $\epsilon$, the equation to be solved (\ref{eq:eqgivZ0}) becomes
\begin{equation}
\label{eq:eqgivZ0fo}
D^2 Z_0^2 + (6D^2+4 \epsilon) Z_0 + 9 D^2 - 4 = 0.
\end{equation}

Since $\bar{Z}$ is the solution of this equation when $\epsilon=0$, we search for a solution of the form $Z_0=\bar{Z}+\eta$, with $\eta \ll 1$. Substituting this value of $Z_0$ in (\ref{eq:eqgivZ0fo}) and limiting ourselves to first order terms in $\epsilon$ and $\eta$, we find the solution
\begin{equation}
\label{eq:expZ0fo}
Z_0= \bar{Z} \left(1 - \frac{\epsilon}{D} \right).
\end{equation}
This allows to calculate
\begin{equation}
Y_0 = \bar{Y} \left(1 + \frac{D-1}{2D} \epsilon\right)
\end{equation}
and
\begin{equation}
F_+(Y_0)= \bar{F} \left(1 - \frac{\epsilon}{6D} \right).
\end{equation}

Inserting these values in the expression (\ref{eq:eigenfun}) and (\ref{eq:altexpressE}) for the energy, we obtain the final result
\begin{equation}
\label{eq:enerfunlm}
	E^{{\rm lm}}_{{\rm f}}=	\sqrt{\frac{\sigma N\ - b}{a}} \left(2a+\frac{\sigma m^2}{2N}\right).
\end{equation}
Here again, it is amazing that the eigenvalues of a so sophisticated Hamiltonian take such a simple form. The contribution $\Delta (m)$ defined by (\ref{mcontur}) can be computed with the value $\bar \nu_0$ derived from (\ref{YUnu}), (\ref{eq:defD}), (\ref{eq:Y0D}). One finds
\begin{equation}
\Delta (m) = \frac{\sigma\, m^2}{2 N} \sqrt{\frac{\sigma\, N-b}{a}}.
\end{equation}
Taking into account (\ref{funnelur}), this is in perfect agreement with (\ref{eq:enerfunlm}).

\section{Yukawa potential}\label{yukpot}
\subsection{Energy spectrum}
To our knowledge, only a few analytical results exist concerning a relativistic Yukawa problem \cite{yuksr,yuksr2}. These results are moreover obtained by solving one-dimensional Hamiltonians, either Klein-Gordon or Dirac, but not a three-dimensional spinless Salpeter Hamiltonian. As we show in this section, the AFM allows to gain some insight about the latter case.  

Let us start from (\ref{ssh}) with $V(r)=-\alpha\, {\rm e}^{-\beta\, r}/r$. With the change of variables $\bm x=\beta\, \bm r$, $\bm q=\bm p/\beta$, one obtains
\begin{equation}\label{hamyuk}
	{\cal H}=\frac{H}{\sigma\beta}=\sqrt{\bm q^{\, 2}+\chi^2}-g\, \frac{{\rm e}^{-x}}{x},
\end{equation}
with 
\begin{equation}\label{chgpar}
	\chi=\frac{m}{\beta},\quad g=\frac{\alpha}{\sigma}.
\end{equation}
We can moreover assume from section~\ref{coulsr} that the eigenvalues of the Hamiltonian $\sqrt{\bm q^{\, 2}+\theta^2}-g/x$ are exactly known and read $\theta\sqrt{1-g^2/N^2}$. Consequently, we can apply the AFM with $V(x)=-g{\rm e}^{-x}/x$ and $P(x)=-g/x$. A quick calculation leads to 
\begin{equation}\label{nudef}
\nu={\rm e}^{-I(\nu)}\left[I(\nu)+1\right].
\end{equation}
Such an equation can be inverted to obtain \cite{af3}
\begin{equation}\label{idefy}
	I(\nu)=-1-W_{-1}\left(-\frac{\nu}{{\rm e}}\right),
\end{equation}
where $W_{-1}$ is the secondary branch of the Lambert function (or Omega function or product-log). The interested reader will find more informations about $W_{-1}(x)$ in \cite{af3}.

Thanks to~(\ref{en1}) and (\ref{idefy}), the eigenenergies of (\ref{hamyuk}) can be written as
\begin{equation}\label{eny}
	E_y(\nu,g)= \chi \sqrt{1-\frac{g^2\nu^2}{N^2}} - \frac{g \nu}{W_{-1}(-\nu/{\rm e})}.
\end{equation}
The physical value $\nu_0$ for $\nu$ is the one ensuring the condition $\left.\partial_\nu E_y(\nu,g)\right|_{\nu=\nu_0}=0$; explicitly, one finds
\begin{equation}\label{eqpary}
1+W_{-1}\left(-\frac{\nu_0}{{\rm e}}\right)+\frac{N^2}{\chi\, \nu_0\, g}\sqrt{1-\frac{g^2\nu_0^2}{N^2}}=0.
\end{equation}
Notice that the property $W'(z)=W(z)/\left[z(1+W(z))\right]$ is used to get this last relation.
The insertion of this value $\nu_0$ into (\ref{eny}) provides the approximate AFM energy
\begin{equation}\label{eny0}
	E_y(\nu_0,g)=\frac{\chi\, N^2+\chi^2\, g\, \nu_0 \sqrt{1-\frac{g^2\nu_0^2}{N^2}}}{\chi\, g\, \nu_0+N^2\sqrt{1-\frac{g^2\nu_0^2}{N^2}}}.
\end{equation}
The final energy spectrum in physical units is given by $\beta\, \sigma\, E_y(\nu_0,g=\alpha/\sigma)$ with $E_y(\nu_0,g)$ given by (\ref{eny0}) and $\nu_0$ solution of (\ref{eqpary}).

Unfortunately, (\ref{eqpary}) is transcendental so that $\nu_0$ cannot be analytically extracted; this drawback is also present in the nonrelativistic version of this problem \cite{af3}. We nevertheless notice that in the limit $\beta\to 0$ one has $\nu_0=1$ and $\beta\, \sigma\, E_y(1,g)$ reduces to (\ref{encoulsr}) as expected.

It is possible to get an asymptotic expression for (\ref{eny0}) in the limit $\chi \to \infty$. In order to simplify the notation, let us set $\epsilon=N/\chi$ a small quantity and $\rho = g/N$. At the limit $\epsilon \to 0$, the third term of (\ref{eqpary}) vanishes and it is easy to check that $\nu_0 \to 1$ by lower values. Thus, it is natural to set $\nu_0 = 1 - \eta$, with $\eta$ a small positive quantity. Then, $1+W_{-1}(-\nu_0/{\rm e}) \to -\sqrt{2 \eta}$ and the third term of (\ref{eqpary}) tends to $\epsilon \sqrt{1-\rho^2}/\rho$ (neglecting the term proportional to $\eta \epsilon$). With these considerations, (\ref{eqpary}) reduces to $\sqrt{2 \eta}=\epsilon \sqrt{1-\rho^2}/\rho$, which immediately provides us with the $\eta$ value and hence the approximate $\nu^*_0$ value. Explicitly, we have
\begin{equation}
\nu^*_0= 1- \frac{N^4}{2\chi^2 g^2}\left(1-\frac{g^2}{N^2}\right).
\end{equation}
One deduces the approximate expression for the energy, valid for large values of $\chi$
\begin{equation}
E_{y}(\nu^*_0,\chi,g)\approx g+\left(\chi-\frac{N^2}{2 \chi}\right)\sqrt{1-\frac{g^2}{N^2}}.
\end{equation}
In physical units (see (\ref{hamyuk}) and (\ref{chgpar})), this last formula reads $E_y=\sigma \beta E_{y}(\nu^*_0,\chi=m/\beta, g=\alpha/\sigma)$, namely
\begin{equation}
 E_y \approx \alpha\beta+
\left(\sigma\, m-\frac{\sigma N^2 \beta^2}{2 \, m}\right)\sqrt{1-\frac{\alpha^2}{\sigma^2 N^2}}.
\end{equation}

\subsection{Critical heights}

It is well known in nonrelativistic quantum mechanics that the Yukawa potential $-g {\rm e}^{-x}/x$ admits a finite number of bound states. Consequently, there exists critical heights, denoted as $g_{nl}$. They are such that if $g>g_{n_0l_0}$, a bound state exists with radial and orbital quantum numbers $n=n_0$ and $l=l_0$ respectively. These critical heights have been intensively studied in the nonrelativistic case \cite{yukgnl,yukgnl2}, but their computation remains problematic with a spinless Salpeter Hamiltonian \cite{yuksrr}. 

Actually, one has to deal with two critical heights in a semirelativistic problem. To understand this fact, let us start from $g=0$, for which $E_y(\nu,\chi,0)=\chi$ logically, and increase its value. Bound states progressively appear; in particular the bound state with quantum numbers $n$ and $l$ appears for $g_{\chi;nl}>0$ such that 
\begin{equation}\label{gc1}
E_y(\nu_{\chi;0},\chi,g_{\chi;nl})=\chi\Rightarrow \nu_{\chi;0}\, g_{\chi;nl}=\frac{N^2}{\chi}.	
\end{equation}
If $g>g_{\chi;nl}$ indeed, $E_y$ will be smaller than the rest mass term $\chi$. Such a negative binding energy shows the existence of a bound state. Replacing $\nu_0$ by $\nu_{\chi;0}$ in (\ref{eqpary}) allows to obtain an explicit expression for $g_{\chi;nl}$, that is
\begin{equation}\label{gc2}
	g_{\chi;nl}=\frac{N^2}{\chi}\, \frac{{\rm e}^{\sqrt{1-\frac{N^2}{\chi^2}}}}{1+\sqrt{1-\frac{N^2}{\chi^2}}}.
\end{equation}
Note that the property $W_{-1}(x\, {\rm e}^x)=x$, which is actually a definition for the Lambert function, has to be used. In the nonrelativistic limit, $\chi\to \infty$ and $\chi g_{\chi;nl}\to {\rm e}N^2/2$ as expected from \cite{af3}, where the nonrelativistic Yukawa problem is solved with the AFM. The factor $1/(2\chi)$ is indeed a consequence of the scaling laws in the nonrelativistic case, while the dependence in $N^2$ has been checked by comparison to numerical data \cite{af3}. 

We now continue to increase $g$. Since (\ref{eny0}) is actually a mass formula, one cannot have a value of $g$ so high that $E_y(\nu_0,\chi,g)<0$. But the numerator of (\ref{eny0}) can never vanish since $\chi$, $\nu$, $N$ and $g$ are positive. The only trivial possibility is $\chi=0$, but it leads to an unphysical state with $E_y=0$. However, there is an obvious constraint stating that $g\nu_0/N\leq 1$. Since $\nu_0\in[0,1]$ from (\ref{nudef}), one has 
\begin{equation}
	g< N.
\end{equation}
Thus the energy (\ref{eny0}) decreases with increasing $g$ until $g=N$ is reached. Then $\nu_0=1$ is a solution of (\ref{eqpary}) and one has the minimal value $E_y(1,\chi,N)=N$. No bound state with $E_y<N$ can exist. 

There are in consequence several constraints ruling the existence of bound states in the Yukawa potential. In order to have a complete spectrum, one must satisfy $\chi> N$ from (\ref{gc2}), and $g_{\chi;nl}<g<\left.N\right|_{n=l=0}$. However, the inequality $g_{\chi;nl}<\left.N\right|_{n=l=0}$ is not trivially satisfied. One has $g_{\chi;nl}/N=f(N/\chi)$, with $f(y)$ an monotonous increasing function of $y\in[0,1]$ such that $f(0)=0$ and $f(1)=1$. Thus, the inequality $N/\chi<f^{-1}(\left.N\right|_{n=l=0}/N)$ should be verified. Finally, the constraints, expressed in physical units, are
\begin{equation}
	m>\beta\ \frac{N}{f^{-1}(\left.N\right|_{n=l=0}/N)},\quad g_{\chi;nl}<\frac{\alpha}{\sigma}<\left.N\right|_{n=l=0}. 
\end{equation}
The existence of the ground state is guaranteed if $m>\beta \left.N\right|_{n=l=0}$ and $g_{\chi;00}<\alpha/\sigma<\left.N\right|_{n=l=0}$.

\section{Improved formulas}\label{increase}

The AFM provides numerous analytical approximate formulas for eigenvalues of the semirelativistic Hamiltonian~(\ref{ham0}). They all depend on a global quantum number $N$. The AFM cannot give clear indications about the exact form of $N$. In previous papers \cite{af,af2,af3,afsqrt}, it is shown that the accuracy of the formula can be greatly improved by fitting the form of $N$ with the exact results. In this section, the same procedure will be implemented for some Hamiltonians considered above. Details are not given here but can be found in \cite{af,af2}. It is worth mentioning that very accurate eigenvalues can be obtained numerically with the Lagrange mesh method \cite{lag}. It is very accurate and easy to implement, even in the case of a relativistic kinematics.

\subsection{Ultrarelativistic power-law potential}\label{numform}

Let us first consider the following dimensionless Hamiltonian 
\begin{equation}
\label{hpowdimless}
H=2\sqrt{\bm q^2}+ x^\lambda
\end{equation}
with $\lambda>0$. The approximate energy spectrum is given by (\ref{ezerom}) with $a=1$ and $\sigma=2$.
With the choice $N=N_2=2\, n+l+3/2$, upper bounds are obtained. As mentioned before, another choice for the $n$- and $l$-dependences of $N$ can greatly improve the results.
By using the form
\begin{equation}
\label{gnlambdapow}
N=b(\lambda)\, n + d(\lambda) \, l + c(\lambda),
\end{equation}
we find smooth variations for coefficients $b$, $c$ and $d$ for $\lambda \in ]0,2]$ ($l \le 3$ and $n \le 3$). In particular, $d(\lambda)\approx 1$ with relative variations less than 2\%. Coefficients $b(\lambda)$ and $c(\lambda)$ can be fitted with various functions and similar agreement. Finally, we choose
\begin{equation}
\label{bcdlambdapow}
b(\lambda)=\frac{3.00\, \lambda+3.67}{\lambda+3.40}, \quad
c(\lambda)=\frac{2.69\, \lambda+8.69}{\lambda+8.27}, \quad
d(\lambda)=1.
\end{equation}
Agreement with exact results is very good but the variational character of the approximation is no longer guaranteed.
With the choice (\ref{bcdlambdapow}), the maximal relative error for $l \le 3$ and $n \le 3$ and for $\lambda \in [0.1,2]$ is located between 0.3 and 1.1\%. With the choice $N=N_2$, the corresponding error is located between 4.5 and 12.7\%. Results for $\lambda =1$ are presented in table~\ref{tab:powla1}.

\begin{table}[htb]
\caption{\label{tab:powla1}Eigenvalues $\epsilon(n,l)$ of the Hamiltonian~(\ref{hpowdimless}) with
$\lambda=1$, for some sets $(n,l)$. 
First line: value from numerical integration; 
second line: approximate result with $N$ defined by (\ref{gnlambdapow}) and (\ref{bcdlambdapow});
third line:  approximate result with $N=2\, n+l+3/2$.}
\begin{indented}
\item[]\begin{tabular}{@{}ccccc}
\br
$l$ & $\epsilon(0,l)$ & $\epsilon(1,l)$ & $\epsilon(2,l)$ & $\epsilon(3,l)$ \\
\mr
0 & 3.1577 & 4.7109 & 5.8913 & 6.8742 \\
  & 3.1338 & 4.6849 & 5.8374 & 6.7973 \\
  & 3.4641 & 5.2915 & 6.6333 & 7.7460 \\

1 & 4.2248 & 5.4575 & 6.4837 & 7.3767 \\
  & 4.2215 & 5.4725 & 6.4866 & 7.3623 \\
  & 4.4721 & 6.0000 & 7.2111 & 8.2462 \\

2 & 5.0789 & 6.1304 & 7.0470 & 7.8671 \\
  & 5.0814 & 6.1602 & 7.0764 & 7.8869 \\
  & 5.2915 & 6.6333 & 7.7460 & 8.7178 \\

3 & 5.8108 & 6.7425 & 7.5775 & 8.3387 \\
  & 5.8156 & 6.7785 & 7.6207 & 8.3787 \\
  & 6.0000 & 7.2111 & 8.2462 & 9.1652 \\
\br
\end{tabular}
\end{indented}
\end{table}

In section~\ref{srho}, it is shown that $\left. N\right|_{l=0}$ must be given by (\ref{N2def}) for a quadratic potential. Formulas~(\ref{bcdlambdapow}) give $b(2)=1.79$ close to $\pi/\sqrt{3}\approx 1.81$ and $c(2)=1.37$ close to $\pi\sqrt{3}/4\approx 1.36$. This is also in agreement with formulas~(71) in \cite{af} which predict, in the case of a nonrelativistic Hamiltonian with a linear potential, $b(1)=1.79$ and $c(1)=1.38$.

\subsection{Relativistic Coulomb potential}
\label{sec:rcoulp}

With dimensionless variables, the semirelativistic Coulomb Hamiltonian is written
\begin{equation}
\label{hcouldimless}
H=2\sqrt{\bm q^2+1}-\frac{a}{x}
\end{equation}
with $a_c \le a < 0$. The approximate energy spectrum is given by (\ref{encoulsr}) with $m=1$ and $\sigma=2$.
With the choice $N=N_{-1}=n+l+1$, upper bounds are obtained and the results are exact in the limit $a\to 0$. As shown in the previous section, another choice for the $n$- and $l$-dependences of $N$ can greatly improve the results.
By using the form
\begin{equation}
\label{gnacoul}
N=b(a)\, n + d(a) \, l + c(a)
\end{equation}
and imposing $b(0)=c(0)=d(0)=1$,
we find smooth variations for coefficients $b$, $c$ and $d$ for $a \in ]0,a_c]$ ($l \le 3$ and $n \le 3$). These coefficients can be fitted with various functions and similar agreement. Finally, we choose
\begin{equation}
\label{bcdacoul}
\fl
b(a)=\frac{1.03\, a-1.48}{a-1.48}, \quad
c(a)=\frac{1.07\, a-1.64}{a-1.64}, \quad
d(a)=\frac{0.96\, a-1.56}{a-1.56}.
\end{equation}
Agreement with exact results is very good for $a\lesssim 1.2$ but the variational character of the approximation is no longer guaranteed. 
With the choice (\ref{bcdacoul}), the maximal relative error for $l \le 3$ and $n \le 3$ and for $a \in [0.2,1.2]$ is located between 0.005 and 0.3\%. With the choice $N=N_{-1}$, the corresponding error is located between 0.004 and 17.3\%. 
To obtain a good accuracy in the domain $a\approx a_c=4/\pi\approx 1.273$, special method must be used as the one presented in \cite{srcoul4}. To recover the value obtained for the ground state in this paper, it is necessary to have $c(a_c)=0.73$. Our formula gives $c(a_c)=0.77$. Results for $a=1$ are presented in table~\ref{tab:coula6}. 

One can see that $N=N_{-1}=n+l+1$ is a better choice for large values of $n$ or $l$. This is can be understood as a kind of nonrelativistic behavior since the limits $a\to 0$ and $N\to \infty$ have similar effects. With $f$ a generic name for the coefficients $b$, $c$ and $d$, a more efficient form for these parameters should be $f(a,n,l)$ with $\lim_{a\to 0}f(a,n,l)=1$, $\lim_{n\to \infty}f(a,n,l)=1$ and $\lim_{l\to \infty}f(a,n,l)=1$. Such a refinement is beyond the scope of the present paper.

\begin{table}[htb]
\caption{\label{tab:coula6}Eigenvalues $\epsilon(n,l)$ of the Hamiltonian~(\ref{hcouldimless}) with
$a=1$, for some sets $(n,l)$. 
First line: value from numerical integration; 
second line: approximate result with $N$ defined by (\ref{gnacoul}) and (\ref{bcdacoul});
third line:  approximate result with $N=n+l+1$.}
\begin{indented}
\item[]\begin{tabular}{@{}ccccc}
\br
$l$ & $\epsilon(0,l)$ & $\epsilon(1,l)$ & $\epsilon(2,l)$ & $\epsilon(3,l)$ \\
\mr
0 & 1.65817 & 1.92184 & 1.96739 & 1.98231 \\
  & 1.65982 & 1.92356 & 1.96680 & 1.98151 \\
  & 1.73205 & 1.93649 & 1.97203	& 1.98431 \\

1 & 1.93515 & 1.97122 & 1.98389 & 1.98973 \\
  & 1.93476 & 1.97012 & 1.98291 & 1.98895 \\
  & 1.93649 & 1.97203 & 1.98431 & 1.98997 \\

2 & 1.97187 & 1.98416 & 1.98987 & 1.99297 \\
  & 1.97296 & 1.98416 & 1.98961 & 1.99266 \\
  & 1.97203 & 1.98431 & 1.98997 & 1.99304 \\

3 & 1.98428 & 1.98993 & 1.99301 & 1.99487 \\
  & 1.98528 & 1.99021 & 1.99302 & 1.99477 \\
  & 1.98431 & 1.98997 & 1.99304 & 1.99489 \\
\br
\end{tabular}
\end{indented}
\end{table}

\subsection{Ultrarelativistic funnel potential}

Written with dimensionless variables, the ultrarelativistic Hamiltonian with the funnel potential is given by
\begin{equation}
\label{hfundimless}
H=2\sqrt{\bm q^2}+x-\frac{\beta}{x}
\end{equation}
with $\beta \ge 0$. The approximate energy spectrum is given by (\ref{funnelur}) with $\sigma=2$, $a=1$ and $b$ replaced by $\beta$ (to avoid confusion with the coefficient of $n$). This kind of Hamiltonian is often used in hadronic physics with typical values for $\beta \approx 0.4$.
With the choice $N=N_2=2\,n+l+3/2$, upper bounds are obtained. As shown in previous sections, another choice for the $n$- and $l$-dependences of $N$ can greatly improve the results.
By using the form
\begin{equation}
\label{gnbetfun}
N=b(\beta)\, n + d(\beta) \, l + c(\beta)
\end{equation}
we find smooth variations for coefficients $b$, $c$ and $d$ for $\beta \in [0,1]$ ($l \le 3$ and $n \le 3$); notice that $\beta$ must be lower than the critical value $4/\pi\approx 1.27$. These coefficients can be fitted with various functions and similar agreement. Finally, we choose
\begin{equation}
\label{bcdbetfun}
\fl
b(\beta)=\frac{1.88\, \beta-5.34}{\beta-3.51}, \quad
c(\beta)=\frac{1.99\, \beta-4.40}{\beta-3.49}, \quad
d(\beta)=\frac{0.76\, \beta-2.46}{\beta-2.54}.
\end{equation}
Agreement with exact results is very good but the variational character of the approximation is no longer guaranteed. With the choice (\ref{bcdbetfun}), the maximal relative error for $l \le 3$ and $n \le 3$ and for $\beta \in [0,1]$ is located between 0.6 and 4.9\%. With the choice $N=N_2$, the corresponding error is located between 12.7 and 42.2\%. 
Results for $\beta =0.4$ are presented in table~\ref{tab:funbet04}.

For $\beta=0$, one obtains $b=1.52$, $c=1.26$, and $d=1.09$. From (\ref{bcdlambdapow}) with $\lambda=1$, one obtains $b=1.52$, $c=1.23$, and $d=1$. These values are close to each other as expected.

\begin{table}[htb]
\caption{\label{tab:funbet04}Eigenvalues $\epsilon(n,l)$ of the Hamiltonian~(\ref{hfundimless}) with
$\beta=0.4$, for some sets $(n,l)$. 
First line: value from numerical integration; 
second line: approximate result with $N$ defined by (\ref{gnbetfun}) and (\ref{bcdbetfun});
third line:  approximate result with $N=2\, n+l+3/2$.}
\begin{indented}
\item[]\begin{tabular}{@{}ccccc}
\br
$l$ & $\epsilon(0,l)$ & $\epsilon(1,l)$ & $\epsilon(2,l)$ & $\epsilon(3,l)$ \\
\mr
0 & 2.7821 & 4.3709 & 5.5874 & 6.5938 \\
  & 2.7804 & 4.4196 & 5.5977 & 6.5678 \\
  & 3.2249 & 5.1381 & 6.5115 & 7.6420 \\

1 & 3.9944 & 5.2365 & 6.2744 & 7.1772 \\
  & 3.9737 & 5.2529 & 6.2765 & 7.1552 \\
  & 4.2895 & 5.8652 & 7.0993 & 8.1486 \\

2 & 4.8993 & 5.9549 & 6.8772 & 7.7028 \\
  & 4.8837 & 5.9710 & 6.8887 & 7.6978 \\
  & 5.1381 & 6.5115 & 7.6420 & 8.6255 \\

3 & 5.6588 & 6.5927 & 7.4311 & 8.1957 \\
  & 5.6489 & 6.6115 & 7.4508 & 8.2046 \\
  & 5.8652 & 7.0993 & 8.1486 & 9.0774 \\
\br
\end{tabular}
\end{indented}
\end{table}

\section{Summary of the results}\label{conclu}

The auxiliary field method, which is strongly connected with the envelope theory \cite{afmenv}, is a powerful tool to compute approximate analytical solutions of the Schr\"odinger equation \cite{af,af2}. This method was already used to deal with semirelativistic Hamiltonians in previous works \cite{afhyb,sema04,srcoul3}. In the present paper, we extend this technique and apply it to various potentials presenting an interest in atomic and hadronic physics: power-law interactions (with special focus on quadratic, linear, and Coulomb potentials), square root potential, funnel potential, and Yukawa interaction. Both nonrelativistic and ultrarelativistic limits are analyzed. Previous results are recovered \cite{env,env2,srcoul3}, but new ones are presented. In particular, closed formulas are computed for the first time. It is shown that their accuracy can be largely improved by slight modifications given by a comparison with exact results coming from a numerical analysis. 

It is worth saying that in hadronic and atomic physics, one is sometimes led to use
running masses, that is basically to deal with a Hamiltonian of the form
$H=\sigma\sqrt{\bm p^2+m^2(r)}+g\, U(r)$, where the mass is
position-dependent \cite{brau02}. Provided that an analytical
solution, $e(m,a)$, of the Hamiltonian $\bm p^2/m+a\, U(r)$ is known, one
can use the AFM with $P(r)=U(r)$ and $V(r)=m^2(r)$. The energy formula
reads $\nu\, E(\nu,\rho)=\sigma^2+m^2(I(\rho))-\rho\, U(I(\rho))+\nu\,
e(\nu,g+\rho/\nu)$, and analytical solutions can then be hoped after
elimination of the auxiliary fields for some particular forms of $m(r)$.
We hope to present explicit applications of this result in future works.

\section*{Acknowledgments}
CS and FB would thank the F.R.S.-FNRS for financial support. The authors thank Fabian Brau for useful discussions.

\begin{appendix}
\section{Unequal masses}\label{uneqma}

It is possible that some particular problems require to deal with a system of two particles with unequal masses. In this case, a general spinless Salpeter Hamiltonian is given in the rest frame by
\begin{equation}
\label{Hm1m2}
	H=\sqrt{\bm p^2+m^2_1}+\sqrt{\bm p^2+m^2_2}+V(\bm r).
\end{equation}

A general but useful result can be obtained for Hamiltonians with two different masses. Let us consider the following two-body Hamiltonians $H=T_1+T_2+V$, $H_1=2 T_1+V$, and $H_2=2 T_2+V$ whose ground state energies are respectively $E=\langle\phi|H|\phi\rangle$, $E_1=\langle\phi_1|H_1|\phi_1\rangle$, and $E_2=\langle\phi_2|H_2|\phi_2\rangle$. Since $H=(H_1+H_2)/2$, we can write
\begin{equation}
\langle\phi|H|\phi\rangle= \frac{1}{2} \left( \langle\phi|H_1|\phi\rangle + \langle\phi|H_2|\phi\rangle \right). 
\end{equation}
The Ritz theorem implies that
\begin{equation}
E \ge \frac{1}{2} \left( E_1 + E_2 \right). 
\end{equation}
For particular cases, this approximation can be quite good. In \cite{sema94}, it is shown that $E \approx ( E_1 + E_2 )/2$ for a relativistic Hamiltonian of kind (\ref{Hm1m2}) with $V(r)=a\, r$ and $m_i \ll \sqrt{a}$.

The square roots appearing in the kinetic terms can still be avoided by resorting to the AFM as exposed in section~\ref{srhamproc}. But this time, two auxiliary fields, $\nu_1$ and $\nu_2$, have to be introduced. One is led to the Hamiltonian
\begin{equation}\label{eq:hamuneq}
	\tilde{H}(\nu_1,\nu_2)=\frac{\nu_1+\nu_2}{2}+\frac{m^2_1}{2\nu_1}+\frac{m^2_2}{2\nu_2}+\frac{\bm p^2}{2M(\nu_1,\nu_2)}+V(\bm r),
\end{equation}
with 
\begin{equation}
	M(\nu_1,\nu_2)=\frac{\nu_1\nu_2}{\nu_1+\nu_2}
\end{equation}
playing the role of a reduced mass. Analytical energy formulas can then be found from (\ref{eq:hamuneq}), but the minimization of the energy with respect to the two auxiliary fields leads to very complicated equations. That is why we have chosen to restrict ourselves to the case $m_1=m_2=m$, implying $\nu_1=\nu_2=\nu=2 M$ by symmetry.

A very interesting solvable case corresponds to the situation where the particles interact via a Coulomb potential and only one of them is massless. Thus, we are concerned with the eigenvalues of the spinless Salpeter equation resulting from the Hamiltonian
\begin{equation}
\label{eq:hcoul0}
	H=\sqrt{\bm p^2}+\sqrt{\bm p^2+m^2}-\frac{a}{r}.
\end{equation}
The introduction of two auxiliary fields $\mu=\nu_1$ and $\nu=\nu_2$ in the nonrelativistic Hamiltonian leads to
\begin{equation}\label{hamuneq}
	\tilde{H}(\mu,\nu)=\frac{\mu+\nu}{2}+\frac{m^2}{2\nu}+\frac{\bm p^2}{2M(\mu,\nu)}- \frac{a}{r}.
\end{equation}
The original Hamiltonian (\ref{eq:hcoul0}) is recovered by a proper elimination of $\mu$ and $\nu$.
The eigenenergies of (\ref{hamuneq}) are given by
\begin{equation}
\label{eq:eigen0}
E(\mu,\nu)=-\frac{a^2}{2 N^2} M(\mu,\nu) + \frac{\mu+\nu}{2}+\frac{m^2}{2\nu},
\end{equation}
with $N=N_{-1}=n+l+1$ \emph{a priori}.

It is easier to begin the minimization by calculating the derivative with respect to $\mu$, $\partial E/\partial \mu = 0$, using the property $\partial M/\partial \mu = M^2/\mu^2$. One obtains the minimal value
\begin{equation}
\mu_0(\nu)=\left ( \frac{a}{N} - 1 \right) \nu.
\end{equation}
Reporting in $E(\mu_0(\nu),\nu)=\nu [a/N - a^2/(2 N^2)] + m^2/(2 \nu)$, one is led to the minimization of this function of $\nu$ only. It is easy to find the value $\nu_0$ which minimizes $E$, and the value of the energy $E = E(\mu_0(\nu_0),\nu_0)$ which corresponds to this minimum. Explicitly, one obtains
\begin{equation}
\label{eq:Ecoul0}
E=2 m \sqrt{\frac{a}{2N} \left(1 - \frac{a}{2N} \right)}.
\end{equation}
In this formula, one has in principle $N=n+l+1$, but, as we pointed out several times, it is justified to take a more sophisticated expression in order to get a better accuracy.
It can be seen from (\ref{eq:Ecoul0}) that $\lim_{m\to 0} E =0$, as expected. 

\section{Some polynomial equations}\label{poleq}

Finding analytical energy formulas for the potentials that we study in this work requires an analytical knowledge of the roots of particular cubic
and quartic equations. We sum up these equations in this appendix and put their roots in a form that is as convenient as possible to deal with. Notice that the needed polynomial equations are the same as those appearing in our previous work \cite{af2}.

We begin by the cubic equation ($Y \ge 0$)
\begin{equation}
\label{eq:redcubeq1}
x^3 \pm 3x - 2Y =0,
\end{equation}
for which there exists only one positive root given by
\begin{equation}
\label{eq:rootcubeq}
F_\pm(Y) = \left(Y + \sqrt{Y^2\pm 1} \right)^{1/3} \mp \left(Y + \sqrt{ Y^2\pm1}
\right)^{-1/3}.
\end{equation}
Written in the above form, it seems that $F_{-}(Y)$ is not properly defined for $Y < 1$. But, for this range of $Y$ values, one can show that
\begin{equation}
F_{-}(Y) = 2 \cos\left( \frac{1}{3} \arccos Y \right).
\end{equation}
So $F_{-}(Y)$ is well defined for all positive values of its argument. 
It can be checked that the following approximate forms hold
\begin{equation}
\label{eq:behYcsmall1}
F_+(Y) \approx \frac{2Y}{3} \quad \textrm{if} \quad Y \ll 1,
\end{equation}
\begin{equation}
\label{eq:behYcsmall2}
F_-(Y) \approx \sqrt 3+\frac{Y}{3} \quad \textrm{if} \quad Y \ll 1,
\end{equation}
\begin{equation}
\label{eq:behYcsmall3}
F_\pm(Y) \approx (2Y)^{1/3} \quad \textrm{if} \quad Y \gg 1.
\end{equation}

The quartic equation which gives the most pleasant form for the roots is ($Y \ge 0$)
\begin{equation}
\label{eq:redcubeq2}
4 x^4 \pm 8x - 3Y =0.
\end{equation}
For each sign, there exists only one positive root given by
\begin{equation}
\label{eq:rootquarteq}
G_{\pm}(Y) = \mp \frac{1}{2} \sqrt{V(Y)} + \frac{1}{2} \sqrt{ 4 (V(Y))^{-1/2}
- V(Y)},
\end{equation}
with
\begin{equation}
\label{eq:defVY}
V(Y)=\left(2 + \sqrt{4 + Y^3} \right)^{1/3} -  Y\left(2 + \sqrt{4 + Y^3}
\right)^{-1/3}.
\end{equation}
The following approximate expressions can also be used for simplicity
\begin{equation}
\label{eq:behYqsmall1}
G_{+}(Y) \approx \frac{3Y}{8} \quad \textrm{if} \quad Y \ll 1,
\end{equation}
\begin{equation}
\label{eq:behYqsmall2}
G_{-}(Y) \approx 2^{1/3} + \frac{Y}{8} \quad \textrm{if} \quad Y \ll 1,
\end{equation}
\begin{equation}
\label{eq:behYqsmall3}
G_{\pm}(Y) \approx \left( \frac{3Y}{4}\right)^{1/4} \quad \textrm{if} \quad Y \gg 1.
\end{equation} 

\end{appendix}


\section*{References}

\begin{thebibliography}{99}

\bibitem{flu} Fl\"{u}gge S 1999 \textit{Practical Quantum Mechanics} (Springer, Berlin) and references therein
\bibitem{af} Silvestre-Brac B, Semay C and Buisseret F 2008 \textit{J. Phys. A: Math. Theor.} \textbf{41} 275301 (arXiv:0802.3601)
\bibitem{af2} Silvestre-Brac B, Semay C and Buisseret F 2008 \textit{J. Phys. A: Math. Theor.} \textbf{41} 425301 (arXiv:0806.2020)
\bibitem{af1} Dirac P A M 1966 \textit{Lectures on Quantum Mechanics} (Belter Graduate School of Sciences, Yeshiva University, New York);  Brink L, Di Vecchia P and Howe P S 1977 \textit{Nucl.\ Phys.\ B} \textbf{118} 76 
\bibitem{af3} Silvestre-Brac B, Semay C and Buisseret F 2008 (arXiv:0811.0287)
\bibitem{hybri} Semay C, Buisseret F and Silvestre-Brac B 2008 (arXiv:0812.3291)
\bibitem{afsqrt} Silvestre-Brac B, Semay C and Buisseret F 2008 (arXiv:0901.4614)
\bibitem{ssh} Salpeter E E and Bethe H A 1951 \textit{Phys. Rev.} \textbf{84} 1232 
\bibitem{ssh2} Salpeter E E 1952 \textit{Phys. Rev.} \textbf{87} 328
\bibitem{env0} Hall R L 1983 \textit{J. Math. Phys.} \textbf{24} 324; 1984 \textit{J. Math. Phys.} \textbf{25} 2708
\bibitem{Hall01} Hall R L, Lucha W and Sch\"{o}berl F F 2001 \textit{J.\ Math.\ Phys.} {\bf 42} 5228 (\textit{Preprint} hep-th/0101223)
\bibitem{env} Hall R L, Lucha W and Sch\"{o}berl F F 2002 \textit{Int. J. Mod. Phys. A} \textbf{17} 1931 (\textit{Preprint} hep-th/0110220)
\bibitem{env3} Hall R L, Lucha W and Sch\"{o}berl F F 2002 \textit{J. Math. Phys.} \textbf{43} 5913 (\textit{Preprint} math-ph/0208042)
\bibitem{env2} Hall R L, Lucha W and Sch\"{o}berl F F 2003 \textit{Int. J. Mod. Phys. A} \textbf{18} 2657 (\textit{Preprint} hep-th/0210149) and references therein
\bibitem{env4} Hall R L and Lucha W 2005 \textit{J. Phys. A} \textbf{38} 7997 (\textit{Preprint} math-ph/0508009)
\bibitem{afmenv} Buisseret F, Semay C and Silvestre-Brac B 2008 (arXiv:0811.0748) to appear in J. Math. Phys.
\bibitem{lucha} Lucha W, Sch\"{o}berl F F and Gromes D 1991 \textit{Phys. Rept.} \textbf{200} 127
\bibitem{bali} Bali G S 2001 \textit{Phys. Rept.} \textbf{343} 1 (\textit{Preprint} hep-ph/0001312)
\bibitem{afhyb} Buisseret F and Mathieu V 2006 \textit{Eur. Phys. J. A} \textbf{29} 343 (\textit{Preprint} hep-ph/0607083)
\bibitem{sema04} Semay C, Silvestre-Brac B and Narodetskii I 2004 \textit{Phys. Rev. D} {\bf 69} 014003 (\textit{preprint} hep-ph/0309256)
\bibitem{feyn} Feynman R P 1939 \textit{Phys. Rev.} \textbf{56} 340;   
Lichtenberg D B 1989 \textit{Phys. Rev. D} \textbf{40} 4196
\bibitem{macd33} MacDonald J K L 1933 \textit{Phys. Rev.} \textbf{43} 830
\bibitem{lucha2} Lucha W and Sch\"{o}berl F F 1995 \textit{Phys. Rev. A} \textbf{51} 4419
\bibitem{srcoul1} Herbst I W 1977 \textit{Commun. Math. Phys.} \textbf{53} 285
\bibitem{srcoul2} Durand B and Durand L 1983 \textit{Phys. Rev. D} \textbf{28} 396; 1994 \textit{Phys. Rev. D} \textbf{50} 6662
\bibitem{srcoul3} Lucha W and Sch\"{o}berl F F 1994 \textit{Phys. Rev. D} \textbf{50} 5443 (\textit{Preprint} hep-ph/9406312); 1996 \textit{Phys. Rev. A} \textbf{54} 3790 (\textit{Preprint} hep-ph/9603429) 
\bibitem{srcoul4} Lucha W and Sch\"{o}berl F F 1996 \textit{Phys. Lett. B} \textbf{387} 573 (\textit{Preprint} hep-ph/9607249)
\bibitem{luchaho} Li Z F, Liu J J, Lucha W, Ma W G and Sch\"{o}berl F F 2005 \textit{J. Math. Phys.} \textbf{46} 103514 (\textit{Preprint} hep-ph/0501268)
\bibitem{Abra} Abramowitz M and Stegun I A 1970 \textit{Handbook of mathematical functions} (New York: Dover publications)
\bibitem{hyb} Allen T J, Olsson M G and Veseli S 1998 \textit{Phys.\ Lett.\  B} {\bf 434} 110 (\textit{Preprint} hep-ph/9804452); Buisseret F and Semay C 2006 \textit{Phys. Rev. D} \textbf{74} 114018 (\textit{Preprint} hep-ph/0610132); Buisseret F, Semay C, Mathieu V and Silvestre-Brac B 2007 \textit{Eur.\ Phys.\ J.\  A} {\bf 32} 123 (\textit{Preprint} hep-ph/0703020)
\bibitem{boul} Boulanger N, Buisseret F, Mathieu V and Semay C 2008 \textit{Eur. Phys. J. A} {\bf 38} 317 (arXiv:0806.3174) 
\bibitem{yuksr} Hall R L 1986 \textit{J. Phys. A} \textbf{19} 2079; 1999 \textit{Phys. Rev. Lett.} \textbf{83} 468 
\bibitem{yuksr2} Rao N A and Kagali B A 2002 \textit{Physics Letters A} \textbf{296} 192; de Castro A S 2006 \textit{Int.\ J.\ Mod.\ Phys.\  A} {\bf 21} 5141 (\textit{Preprint} hep-th/0507025);  Castro L B and de Castro A S 2007 \textit{Int.\ J.\ Mod.\ Phys.\  E} {\bf 16} 2998 (arXiv:0709.3282)
\bibitem{yukgnl} Hulth\'{e}n L and Laurikainen K V 1951 \textit{Rev. Mod. Phys.} \textbf{23} 1; Green A E S 1982 \textit{Phys. Rev.} A \textbf{26} 1759; Green A E S, Schwartz J M and Khosravi A 1986 \textit{Phys. Rev.} A \textbf{33} 2087
\bibitem{yukgnl2} Brau F and Calogero F 2003 \textit{J. Math. Phys.} \textbf{44} 1554; \textit{J. Phys. A: Math. Gen.} \textbf{36} 12021; Brau F 2003 \textit{J. Phys. A} \textbf{36} 9907 (\textit{Preprint} math-ph/0401023)
\bibitem{yuksrr} Brau F 2005 \textit{J. Nonlin. Math. Phys.} \textbf{12} S86 (\textit{Preprint} math-ph/0411009); \textit{J. Math. Phys.} \textbf{46} 032305 (\textit{Preprint} math-ph/0412042)
\bibitem{lag} Semay C, Baye D, Hesse M and Silvestre-Brac B 2001 \textit{Phys. Rev. E} \textbf{64} 016703 
\bibitem{sema94} Semay C 1994 \textit{J. Phys. G\string: Nucl. Part. Phys.} {\bf 20} 689
\bibitem{brau02} Brau F and Semay C 2002 \textit{J. Phys. G\string: Nucl. Part. Phys.} {\bf 28} 2771
(\textit{Preprint} hep-ph/0412177)

\end{thebibliography}
\end{document}